\documentclass[a4paper,fleqn,10pt]{article}
\pdfoutput=1
\usepackage{amsmath}
\usepackage{amssymb}
\usepackage{array}
\usepackage{calc}
\usepackage{longtable}
\usepackage{multirow}
\usepackage{slashed}
\usepackage{pstricks}
\usepackage{graphicx}
\usepackage{xspace}
\usepackage{units}
\usepackage{tikz}
\usepackage{textcomp}

\numberwithin{equation}{section}
\usepackage{cite}
\usepackage[pdfborder={0 0 0}]{hyperref}
\usepackage[format=hang,labelfont=bf,hypcap=true]{caption}
\usepackage{subcaption}
\usepackage{sectsty}
\usepackage{enumitem}
\usepackage{booktabs}
\usepackage{rotating}
\usepackage{placeins}
\usepackage{cleveref}
\allsectionsfont{\sffamily}
\subsubsectionfont{\mdseries\itshape\large}
\makeatletter
\DeclareRobustCommand*{\bfseries}{%
  \not@math@alphabet\bfseries\mathbf
  \fontseries\bfdefault\selectfont
  \boldmath
}
\makeatother
\let\spreprint\empty
\newcommand{\preprint}[1]{\def\spreprint{\protect#1}}
\let\sinstitute\empty
\newcommand{\institute}[1]{\def\sinstitute{\protect#1}}
\makeatletter
\renewcommand{\maketitle}{\begingroup
  \null\thispagestyle{empty}%
    \ifx\spreprint\empty
      \vskip 5ex
    \else
      \flushright\large\spreprint\vskip 10ex
    \fi
    \vskip 5ex
    \centering
      {\sffamily\bfseries\huge\@title}\vskip 10ex
    \flushleft
      \@author\vskip 2ex
      \ifx\sinstitute\empty
      \else
        {\small\sinstitute}
      \fi
    \vskip 5ex
  \endgroup
}
\makeatother
\renewenvironment{abstract}{\begin{center}
  {\large\sffamily\bfseries Abstract: }
  \begin{minipage}[t]{0.75\textwidth}
}{\end{minipage}\end{center}\vskip 10ex}

\numberwithin{equation}{section}
\allowdisplaybreaks[2]

\newcommand{\PowhegBox}{P\protect\scalebox{0.8}{OWHEG}B\protect\scalebox{0.8}{OX}\xspace}

\newcommand{\Herwig}{H\protect\scalebox{0.8}{ERWIG}\xspace}

\newcommand{\Madgraph}{M\protect\scalebox{0.8}{AD}G\protect\scalebox{0.8}{RAPH}\xspace}

\newcommand{\Pythia}{P\protect\scalebox{0.8}{YTHIA}\xspace}

\newcommand{\Stripper}{S\protect\scalebox{0.8}{TRIPPER}\xspace}

\newcommand{\Mocanlo}{M\protect\scalebox{0.8}{OCANLO}\xspace}
\newcommand{\Phantom}{P\protect\scalebox{0.8}{HANTOM}\xspace}

\newcommand{\Rivet}{R\protect\scalebox{0.8}{IVET}\xspace}

\newcommand{\Recola}{R\protect\scalebox{0.8}{ECOLA}\xspace}

\newcommand{\Sherpa}{S\protect\scalebox{0.8}{HERPA}\xspace}
\newcommand{\Sherpan}{S\protect\scalebox{0.8}{HERPA}}
\newcommand{\Comix}{C\protect\scalebox{0.8}{OMIX}\xspace}

\long\def\symbolfootnote[#1]#2{\begingroup%
\def\thefootnote{\fnsymbol{footnote}}\footnote[#1]{#2}\endgroup}

\newcommand{\order}{\mathcal{O}}

\newcommand{\mc}[1]{\mathcal{#1}}

\newcommand{\bea}{\begin{eqnarray}}
\newcommand{\eea}{\end{eqnarray}}
\newcommand{\bi}{\begin{itemize}}
\newcommand{\ei}{\end{itemize}}

\xspace
\xspace
\xspace
\xspace
\xspace
\xspace
\xspace
\xspace
\xspace
\xspace
\xspace
\xspace
\xspace
\xspace
\xspace
\xspace
\xspace
\xspace
\xspace
\xspace
\xspace
\xspace
\xspace
\xspace
\xspace

\newcommand{\PH}{\ensuremath{h}\xspace}

\newcommand{\Pj}{\ensuremath{j}\xspace}

\newcommand{\Pe}{\ensuremath{e}\xspace}
\newcommand{\Pb}{\ensuremath{b}\xspace}

\newcommand{\Pg}{\ensuremath{g}}
\newcommand{\PW}{\ensuremath{W}\xspace}
\newcommand{\PZ}{\ensuremath{Z}\xspace}
\newcommand{\nnloqcd}{NNLO~QCD\xspace}
\newcommand{\lops}{LO+PS\xspace}

\newcommand{\lo}{LO\xspace}

\newcommand{\gev}{\ensuremath{~\mathrm{GeV}}}

\DeclareMathOperator*{\sumav}{\overline{\sum}}

\DeclareRobustCommand{\loongrightarrow}{%
  \DOTSB\relbar\joinrel\relbar\joinrel\longrightarrow
}
\DeclareRobustCommand{\looongrightarrow}{%
  \DOTSB\relbar\joinrel\relbar\joinrel\loongrightarrow
}

\newlist{myitemize}{itemize}{3}
\setlist[myitemize]{leftmargin=14em}

\newcolumntype{C}{>{\centering\arraybackslash}p{0.14\textwidth}}

\newlength{\unitcharwidth}
\hypersetup{
  pdfauthor={Mareen Hoppe, Marek Schoenherr, Frank Siegert},
  pdftitle={Polarised cross sections for loop-induced processes with \Sherpa+\Recola}
}
\preprint{IPPP/26/72\\MCnet-26-29\\COMETA-2026-41}
\author{Mareen Hoppe${}^{(a)}$, Marek Sch{\"o}nherr${}^{(b)}$, Frank Siegert${}^{(a)}$}
\title{Polarised cross sections for\\[5mm] loop-induced processes with \Sherpan+\Recola}
\institute{${}^{(a)}$~Institut f\"ur Kern- und Teilchenphysik, Technische Universit\"at Dresden, 01062 Dresden, Germany\\
${}^{(b)}$~Institute for Particle Physics Phenomenology, Department of Physics, Durham University, Durham DH1 3LE, United Kingdom\\
$~$\\
Emails:  \href{ }{\color{black} mareen.hoppe@tu-dresden.de}\,, \href{ }{\color{black} marek.schoenherr@durham.ac.uk}\,, \href{ }{\color{black} frank.siegert@tu-dresden.de} }

\begin{document}
\vspace*{5mm}
\maketitle
\vspace*{10mm}
\begin{abstract}
  Studying vector boson polarisation has become an integral part of the LHC physics programme in recent years, as it provides an independent test of the electroweak symmetry-breaking mechanism and the gauge sector of the electroweak Standard Model. These measurements require polarised templates from parton-shower Monte Carlo simulations. Their accuracy does not yet meet the precision requirements of upcoming measurements, leading to significant modelling uncertainties in experimental analyses that will constrain the LHC's polarisation sensitivity at latest in the HL-LHC phase. Polarised loop-induced processes have so far not been included in most parton-shower event generators, despite contributing significantly to processes of interest. In this work, we present the extension of the polarisation framework in the Monte Carlo event generator \Sherpa to polarised loop-induced processes at fixed leading order~(LO) and LO matched to parton shower. In addition to a description of the methodology based on polarised amplitudes from the \Recola matrix element generator, we discuss polarisation effects in relevant inclusive, loop-induced processes and their jet-associated counterparts, several of them for the first time. Both the influence of different polarisation definitions and of QCD and QED resummation effects are investigated. 
\end{abstract}
\newpage
\tableofcontents
\section{Introduction}
\label{sec:intro}
The measurement of vector boson polarisation has become an integral part of the LHC analysis program in recent years~\cite{CMS:2011kaj, ATLAS:2012au, CMS:2015cyj, ATLAS:2016fbc, ATLAS:2016rnf, CMS:2016asd, ATLAS:2019bsc, ATLAS:2019rob, CMS:2020cph, CMS:2020etf, CMS:2020ezf, CMS:2021icx, ATLAS:2022oge, ATLAS:2023zrv,ATLAS:2024qbd, ATLAS:2025wuw,ATLAS:2026kzu,ATLAS:2026fhf}. The direct connection between the longitudinal polarisation of the massive vector bosons and the Higgs mechanism provides a complementary approach to study the Higgs sector, beyond on-shell Higgs boson production processes. Potential new physics effects, most likely to be expected in the electroweak~(EW) sector, may impact different polarisation components unevenly, making polarised observables a sensitive probe for such deviations. 

Since the first polarisation measurements based on Run~1 and early Run~2 data, the measurement strategy has mostly changed from measuring angular coefficients of the vector boson decay products to using template-fits constructed from polarised Monte Carlo events. Polarisation effects based on polarised cross sections have been extensively studied in various processes, including $W^+j$ production up to NNLO QCD~\cite{Pellen:2021vpi}, diboson production up to NNLO QCD and NLO EW~\cite{Denner:2020eck,Denner:2020bcz,Denner:2021csi,Denner:2022riz,Denner:2023ehn, Dao:2023pkl, Dao:2023kwc,Dao:2024ffg, Le:2022lrp, Le:2022ppa, Poncelet:2021jmj, Carrivale:2025mjy, Pelliccioli:2025com}, vector boson scattering up to NLO QCD+EW~\cite{Ballestrero:2017bxn, Ballestrero:2019qoy, Ballestrero:2020qgv, Denner:2024tlu, Denner:2025xdz, Denner:2026vmf} and triboson production \@ LO~\cite{Le:2026oux}. Public tools capable of providing fixed-order calculations are \Madgraph~\cite{Alwall:2014hca, BuarqueFranzosi:2019boy} and \Sherpa~\cite{Sherpa:2024mfk,Sherpa:2019gpd, Hoppe:2023uux} at LO as well as \Recola+\Mocanlo at NLO QCD+EW~\cite{Denner:2026phn}. For more realistic predictions at the hadron level, polarised event generation is possible with \Madgraph interfaced with \Pythia~\cite{Bierlich:2022pfr} or \Herwig~\cite{Bahr:2008pv, Bellm:2015jjp}, and \Sherpa, providing LO and nLO QCD accuracy matched to the parton shower~(PS) for nearly arbitrary processes involving intermediate vector bosons. Polarised inclusive diboson production can furthermore be simulated by \PowhegBox~\cite{Nason:2004rx, Frixione:2007vw, Alioli:2010xd, Pelliccioli:2023zpd} interfaced with \Pythia up to NLO QCD+PS. \Phantom~\cite{Ballestrero:2007xq, Ballestrero:2017bxn} can provide predictions for polarised 2$\rightarrow$6 processes at LO+PS. SMEFT effects in inclusive $W^+Z$ production at NLO+PS were recently studied as well~\cite{Haisch:2025jqr}.   

In most cases, realistic simulation of polarised events remains limited to tree-level Born processes. However, although loop-induced processes are formally of NNLO relative to their quark-induced counterparts, they can contribute significantly e.g. to inclusive diboson production at hadron colliders due to the large gluon content of the proton. Only recently, the \Madgraph collaboration introduced a method to enable their simulation beyond fixed-order setups~\cite{Javurkova:2024bwa}. 

In this paper, we present an extension of \Sherpa's polarisation framework~\cite{Hoppe:2023uux} to enable the simulation of loop-induced processes at fixed LO and LO matched to a parton shower~(PS). The paper is organised as follows: In Section~\ref{sec:methods:algo}, we briefly review the formalism used to define and compute polarised cross sections and describe the newly implemented methods for simulating polarised loop-induced processes. Section~\ref{sec:method:val} validates our implementation by comparing against existing results. In Section~\ref{sec:results}, we demonstrate the capabilities of our new implementation by providing results for common loop-induced processes.    	 

\section{Polarised cross sections}
\label{sec:methods}

In this section we briefly review our formalism to define and
calculate polarised cross sections, before validating our
implementation against known results.

\subsection{Definitions and methodology}
\label{sec:methods:algo}

In order to define the contribution of a polarisation
state of an intermediary vector boson, we recall the
following identity replacing the numerator of a massive
vector boson propagator with the sum over \emph{all}
polarisations,
\begin{equation}
  \begin{split}
	-g^{\mu\nu}+\frac{p^\mu p^\nu}{m^2}
    \,=&\;
      \sum_{\lambda=1}^4\;
      \varepsilon_\lambda^{*\mu}(p)\,\varepsilon_\lambda^{\nu}(p)\;.
  \end{split}
\end{equation}
This sum contains the three physical polarisation states,
two transverse and one longitudinal, as well as a fourth
unphysical polarisation state which vanishes if the propagator
momentum approaches its on-shell value.
With its help we can factorise any matrix-element with a resonant
$s$-channel massive vector boson propagator into a production
and decay process, $\mc{M}^\text{prod}$ and $\mc{M}^\text{dec}$,
respectively, linked through the propagation of one of
these four polarisation states $\varepsilon_\lambda$,
\begin{equation}
  \begin{split}
    \mc{M}
    \,=&\;
	  \mc{M}_\mu^\text{prod}
	  \left(
		\frac{i\left(-g^{\mu\nu}+\frac{p^\mu p^\nu}{m^2}\right)}
		     {p^2-m^2+im\Gamma}
	  \right)
	  \mc{M}_\nu^\text{decay}
	\,=\;
	  \sum_{\lambda=1}^4
	  \frac{
		i\;
		\mc{M}_\lambda^\text{prod}
		\cdot
		\mc{M}_\lambda^\text{decay}
	  }{p^2-m^2+im\Gamma}
	\,\equiv\;
	  \sum_{\lambda=1}^4
	  \frac{i\,\widetilde{\mc{M}}_\lambda}{p^2-m^2+im\Gamma}\;.
  \end{split}
\end{equation}
Their direct product, $\widetilde{\mc{M}}_\lambda$,
is introduced as a short-hand.
Using the narrow-width approximation (NWA) for this resonant
$s$-channel massive vector boson, we can now not only promote
$\mc{M}^\text{prod}$ and $\mc{M}^\text{dec}$ to separately
well-defined amplitudes, containing only on-shell external
particles each, but can also simplify the remaining propagator
according to
\begin{equation}\label{eq:methods:nwa}
  \begin{split}
	\frac{1}{(p^2-m^2)^2+m^2\Gamma^2}
	\quad
	\stackrel{\Gamma\ll m}\looongrightarrow
	\quad
	\frac{\pi}{m \Gamma}\,\delta\left(p^2-m^2\right)\;.
  \end{split}
\end{equation}
Hence, the squared amplitude determining the cross section
of the correlated production-times-decay process, connected
through an on-shell intermediate massive vector boson summed
over all its polarisation states, reads
\begin{equation}
  \begin{split}
    \left|\mc{M}\right|^2
	\quad
	\stackrel{\text{NWA}}{\looongrightarrow}
	\quad&\;
	  \frac{\pi}{m\Gamma}
	  \sum\limits_{\lambda,\lambda'=\pm,0}\,
	  \widetilde{\mc{M}}_\lambda\,
	  \widetilde{\mc{M}}_{\lambda'}\;
	  \delta\left(p^2-m^2\right)
	\,\equiv\;
	  \sum\limits_{\lambda=\pm,0}\,
	  \left|\mc{M}\right|_\lambda^2
	  +
	  \sum\limits_{{\lambda,\lambda'=\pm,0}\atop{\lambda\neq\lambda'}}\!\!
	  \left|\mc{M}\right|_{\lambda\lambda'}^2\;,
  \end{split}
\end{equation}
where the $|\mc{M}|_{\lambda\lambda'}^2=\frac{\pi}{m\Gamma}\,\widetilde{\mc{M}}_\lambda\widetilde{\mc{M}}_{\lambda'}^*\,\delta(p^2-m^2)$
and $|\mc{M}|_\lambda^2\equiv|\mc{M}|_{\lambda\lambda'}^2 \delta_{\lambda\lambda'}$.
As the intermediate vector boson is now on-shell, only
its physical polarisations contribute.
The first term on the right-hand side represents the
classical approximation where the intermediate vector
boson is in a well-defined polarisation state, whereas
the second term constitutes the quantum interference
between different polarisation states.\footnote{
  Often, but not always, the quantum interference of
  different polarisation states can be neglected when
  compared to the intrinsic uncertainty of the method
  of the NWA which is of $\order(\frac{\Gamma}{m})$.}

In practice, multiboson processes are often of particular
interest, and the correlation of their respective polarisation
states is to be calculated.
The above use of the NWA is thus generalised to the case of
multiple resonant intermediate vector bosons
\begin{equation}
  \begin{split}
    \left|\mc{M}\right|_{\lambda_1\ldots\lambda_n,\lambda_1'\ldots\lambda_n'}^2
    \,=&\;
	  \frac{1}{N}\;
	  \left|\mc{M}^\text{prod}\right|_{\lambda_1\ldots\lambda_n,\lambda_1'\ldots\lambda_n'}^2\,
	  \prod\limits_{i=1}^n
	  \left|\mc{M}_i^\text{dec}\right|_{\lambda_i\lambda_i'}^2\;,
  \end{split}
\end{equation}
which forms the basis of the spin-correlation algorithm of
\cite{Collins:1987cp, Knowles:1987cu, Knowles:1988vs, Richardson:2001df} as implemented in \Sherpa \cite{Hoche:2014kca}.
{$N=\prod_{i=1}^n n_i$} is a normalisation constant, each $n_i$
ensuring that the decay matrices
$|\mc{M}_i^\text{dec}|_{\lambda_i\lambda_i'}$ have a unit
trace, a crucial feature to generate the decay kinematics
probabilistically.

While the decay matrices are automatically generated at
LO accuracy within \Sherpa with the help of \Comix'
\cite{Gleisberg:2008fv} helicity amplitudes
and in \Comix' polarisation basis, the loop-induced
polarised production-amplitude tensor
$\left|\mc{M}^\text{prod}\right|_{\lambda_1\ldots\lambda_n,\lambda_1'\ldots\lambda_n'}^2$
is calculated using \Recola \cite{Actis:2012qn, Actis:2016mpe, Biedermann:2017yoi}.
As \Recola employs its own polarisation basis, we have
to perform a basis transformation \cite{Hoppe:2023uux}
before its amplitudes can be fed into the
polarisation-correlation algorithm.
We start by calculating the helicity amplitudes
for the respective $2\to m-2$ scattering process
in \Recola's native colour-flow decomposition and
polarisation basis,
\begin{equation}\label{eq:method:recola_amp}
  \begin{split}
	\mc{A}_{\,\,h_1\,\,\ldots\,\,h_m}^{\binom{i_1}{j_1}\ldots\binom{i_m}{j_m}}
	\,=\;
	  \sum_\pi
	  \delta_{j_{\pi(1)}}^{i_1}\,
	  \delta_{j_{\pi(m)}}^{i_m}\,
	  \mc{A}_{\,\,h_1\,\,\ldots\,\,h_m}^{\pi(1)\ldots\pi(m)}
	  \;,
  \end{split}
\end{equation}
where the $h_i$ are the helicity (spin or polarisation) indices
of particle $i$ and the $\binom{i_i}{j_i}$ are the
colour-anti-colour pairs in the
$3\otimes\overline{3}$ decomposition.
The $\mc{A}_{\,\,h_1\,\,\ldots\,\,h_m}^{\pi(1)\ldots\pi(m)}$
are the structure-dressed amplitudes, and the $\pi(i)$ are
all possible permutations of the anti-colour indices $j_i$.
We then compute the colour-summed and -averaged spin-amplitude tensor
\begin{equation}
  \begin{split}
    \left|\mc{M}^\text{prod}\right|_{h_1h_1'\ldots h_mh_m'}^2
    \,=&\;
      \sumav\limits_{\binom{i_1}{j_1},\ldots,\binom{i_m}{j_m}}
      \mc{A}_{\,\,h_1\,\,\ldots\,\,h_m}^{\binom{i_1}{j_1}\ldots\binom{i_m}{j_m}}
      \mc{A}_{\,\,h_1'\,\,\ldots\,\,h_m'}^{*\,\binom{i_1}{j_1}\ldots\binom{i_m}{j_m}}
  \end{split}
\end{equation}
This tensor is still differential in the helicity
indices of \emph{all} external particles.
As we are interested in the contributions of individual
polarisation states of the final state vector bosons
with indices $\{\beta_1,\ldots,\beta_n\}$ only, we sum
and average over the helicities of all other external
states,
\begin{equation}
  \begin{split}
    \left|\mc{M}^\text{prod}\right|_{\lambda_{\beta_1}\lambda_{\beta_1}'\ldots \lambda_{\beta_n}\lambda_{\beta_n}'}^2
    \,=&\;
	  \frac{1}{4}
      \left|\mc{M}^\text{prod}\right|_{h_1h_1'\ldots h_mh_m'}^2
	  \prod_{i\notin\{\beta_1\,\ldots,\beta_n\}}
	  \delta_{h_ih_i'}\,,
  \end{split}
\end{equation}
summing over all helicity indices that are \emph{not} one of the
polarisation indices $\lambda_i$ we want to keep open.
At this point, the polarisation tensor is built in \Recola's
native polarisation basis, and we now transform it into
\Sherpa's native basis \cite{Gleisberg:2008fv, Dittmaier:1998nn}
using the transformations of Ref.~\cite{Hoppe:2023uux}.
Thereafter, the spin amplitudes can be computed in the
framework of Ref.~\cite{Hoppe:2023uux} for all
intermediate vector bosons in all desired reference frames.

To remedy a few of the short-comings of the NWA, we recover the
dominant kinematic off-shell effects near the resonance
by \textit{a posteriori} adjusting the momenta of the
respective resonant vector boson according to the Breit-Wigner
distribution of their off-shell propagator \cite{Hoche:2014kca}.

Phase-space integration channels to compute the underlying unpolarised production process are generated using AMEGIC~\cite{Krauss:2001iv} based on the corresponding tree-level process.  

The same methodology can also be applied to use loop-induced
production-amplitude tensors build from OpenLoops
\cite{Cascioli:2011va, Buccioni:2019sur} helicity amplitudes.
The required interfaces are currently under development, but
are left to be subject of a future publication.

\subsection{Validation}
\label{sec:method:val}
We validate our basis rotation of \Recola's polarisation
definition into \Sherpa's internal definition using
tree-level matrix elements, and compare
against equivalent calculations with
\Sherpa's internal matrix element generator \Comix, both
on the level of individual phase-space points and on the
level of distributions.
The calculation of the complete polarised loop-induced
cross-sections is validated against existing literature
results for $W^+W^-$ production~\cite{Poncelet:2021jmj},
using the setup defined therein.
Fiducial cross sections are shown in Tab.~\ref{tab:validation_WW},
showing excellent agreement for both the off-shell computation (short "full")
and the individual polarisation fractions.
The unpolarised (short "unpol") and polarised cross sections computed
in our improved narrow-width approximation differ from
the ones calculated in the double-pole approximation of Ref.~\cite{Poncelet:2021jmj} though, but never by more than
the expected intrinsic uncertainty of $\order(\Gamma/m)$.

First results computed with an early incarnation of this
implementation are part of the study in Ref.~\cite{Carrivale:2025mjy},
where predictions of polarised cross sections for quark-
and loop-induced inclusive
$ZZ$ production are compared between all available tools.
Our results are in very good agreement with the other tools
within the accuracy expected due to different approximation
applied.

\begin{table}
	\centering
	\renewcommand{\arraystretch}{1.3}
	\begin{tabular}{crrrrrrrr}
	  \toprule
	  & & full
	    & unpol.
	    & LL
	    & LT
		& TL
		& TT
		& int \\
	  \midrule
	  \Sherpa & XS [fb]
	    & $13.892(8)$
	    & $13.543(5)$
	    & $0.718(1)$
	    & $1.833(1)$
	    & $1.833(1)$
	    & $9.758(4)$
	    & $-0.599(1)$ \\
	  (NWA) & \%
	    &
	    & $100$
	    & $5.3$
	    & $13.5$
	    & $13.5$
	    & $72.1$
	    & $-4.4$ \\
	  \midrule
	  \Stripper & XS [fb]
	    & $13.8(2)\phantom{00}$
	    & $13.28(3)\phantom{0}$
	    & $0.695(2)$
	    & $1.790(5)$
	    & $1.774(5)$
	    & $9.58(2)\phantom{0}$
	    & n.a. \\
	  (DPA) & \%
	    &
	    & $100$
	    & $5.2$
	    & $13.5$
	    & $13.4$
	    & $72.1$
	    & \\
	  \bottomrule
	\end{tabular}
  \caption{
	Fiducial cross sections and polarisation fractions for
	loop-induced $W^+W^-$ production at LO, compared to
	the results using \Stripper in the double-pole approximation
	(DPA) reported in Ref.~\cite{Poncelet:2021jmj}.
	Polarisation states are defined in the laboratory frame.
    \label{tab:validation_WW}
  }
\end{table}

\section{Results}
\label{sec:results}
\subsection{Event generation and analysis setup}
To demonstrate the new implementation, we present polarised results for the following loop-induced processes in an LHC setup with a centre-of-mass energy of 13.6 TeV assuming SM dynamics: 
\begin{itemize}
	\item[-] \makebox[5cm][l]{$gg\to W^+W^- \to e^+ \nu_e \mu^- \bar{\nu}_\mu$} at $\mathcal{O}(\alpha_S^2 \alpha^4)$
	\item[-] \makebox[5cm][l]{$gg\to ZZ \to e^+ e^- \mu^+ \mu^-$} at $\mathcal{O}(\alpha_S^2 \alpha^4)$ 
	\item[-] \makebox[5cm][l]{$gg\to Z\gamma \to \mathrm{e}^+ \mathrm{e}^-$ $\gamma$} at $\mathcal{O}(\alpha_S^2 \alpha^3)$
	\item[-] \makebox[5cm][l]{$gg\to Zh \to \mu^+ \mu^- b\bar{b}$} at $\mathcal{O}(\alpha_S^2 \alpha^4)$ 
	\item[-] \makebox[5cm][l]{$gg\to ZZg \to e^+ e^- \mu^+ \mu^-g$} at $\mathcal{O}(\alpha_S^3 \alpha^4)$
	\item[-] \makebox[5cm][l]{$gg\to W^+W^-g \to e^+ \nu_e \mu^- \bar{\nu}_{\mu}g$} at $\mathcal{O}(\alpha_S^3 \alpha^4)$
	\item[-] \makebox[5cm][l]{$gg\to Z\gamma g \to \mathrm{e}^+ \mathrm{e}^-$ $\gamma g$} at $\mathcal{O}(\alpha_S^3 \alpha^3)$
	\item[-] \makebox[5cm][l]{$gg\to Zhg \to \mu^+ \mu^- b\bar{b}g$} at $\mathcal{O}(\alpha_S^3 \alpha^4)$ .
\end{itemize}
Simulations are performed with SHERPA in three different setups: fixed LO, LO matched to SHERPA's default dipole PS~\cite{Schumann:2007mg} and LO+PS including additional QED radiation obtained by YFS resummation~\cite{Yennie:1961ad, Schonherr:2008av} including the photon-splitting corrections of Ref.~\cite{Flower:2022iew}. They are dubbed LO, LO+CSS(QCD) and LO+CSS(QCD)+YFS(QED) in the following discussions. While being technically possible, we do not consider non-perturbative effects here as their contribution to the polarisation decomposition is expected to be negligible. Results obtained with the NWA are compared with full off-shell calculations. For $Zh$ production, the Higgs boson remains on-shell in all calculations to avoid large QCD backgrounds due to its hadronic decay. 

The factorisation and renormalisation scales are set to the arithmetic mean of the two intermediate bosons if they are massive: 
\begin{equation}
	\mu_F = \mu_R = \frac{M_{V_1}+M_{V_2}}{2}.
\end{equation}
For the $Z\gamma$ process, they are set to 
\begin{equation}
	\mu_F = \mu_R = M_{Z\gamma}. 
\end{equation} 
Table~\ref{Tab.: simulation settings} summarises all simulation parameters that are shared among all processes. All fermions are considered massless except the bottom quark in $Zh$ production and $W^+W^-$ production~\footnote{For jet-associated $W^+W^-g$ production, the bottom quarks are assumed to be massless to stabilise the phase-space-integration.}, where the bottom mass is set to $m_b=4.92~\mathrm{GeV}$.

\begin{table}[t]	
	\renewcommand{\arraystretch}{1.3}
	\centering
	\begin{tabular}{ll}
		\toprule
		Setting & \\
		\midrule
		PDF-Set from LHAPDF6 \cite{Buckley:2014ana}  & NNPDF31\_nnlo\_as\_0118\_luxqed~\cite{Manohar:2016nzj, Manohar:2017eqh, Bertone:2017bme} \\
		Electroweak scheme $G_{\mathrm{\mu}}$~\cite{Denner:2000bj} &  $G_{\mathrm{\mu}}=1.16638\cdot10^{-5}\, \mathrm{GeV}^{-2}$ and $\alpha_{EW}=\frac{\sqrt{2}G_\mu M_W^2}{\pi}\left(1-\frac{M_W^2}{M_Z^2}\right)$\\
		& complex mass scheme~\cite{Denner:1999gp, Denner:2005fg, Denner:2006ic} (full), real EW parameters (polarisation) \\
		Strong coupling $\alpha_S(M_\mathrm{Z})$ & 0.118\\
		Vector boson pol masses &$M_W=80.350~\mathrm{ GeV}$, $M_Z=91.153~\mathrm{ GeV}$ \\
		Vector boson pol widths & zero for all VBs in polarisation calculations, \\
		&$\Gamma_W=2.084~\mathrm{ GeV}$,  $\Gamma_Z=2.4943~\mathrm{ GeV}$ otherwise\\
		Top quark, Higgs boson masses & $M_t=172.69~\mathrm{ GeV}$, $m_H=125.25~\mathrm{ GeV}$ \\
		Top quark, Higgs boson width & $\Gamma_{t}=0~\mathrm{GeV}$, $\Gamma_{H}=0.0041~\mathrm{GeV}$, zero for $Zh$ production\\
		\bottomrule
	\end{tabular}
	\caption{Simulation settings identical for all processes.}
	\label{Tab.: simulation settings}
\end{table}

In order to allow for a direct comparison of our NWA results with results obtained by applying the \mbox{(double-)pole} approximation~((D)PA), we use LO partial widths and PDG decay widths to calculate the branching ratios for the intermediate, weak vector bosons. For the Higgs boson decay, however, the PDG branching ratio is assumed due to the large impact of QCD corrections. The mass smearing is performed based on PDG decay widths. 

The phase spaces are chosen according to the most recent ATLAS analyses~\cite{ATLAS:2019rob, ATLAS:2023zrv, ATLAS:2022wnf, ATLAS:2024yzu}, respectively as summarised in Tab.~\ref{Tab.: Phase space selections}.

\begin{table}[!b]	
	\renewcommand{\arraystretch}{1.3}
	\centering
	\begin{tabular}{ll}
		\toprule
		Process & Phase-space selections \\
		\midrule
		$ZZ$ & $p_{T, \,e^\pm} >7\gev$, $p_{T, \,\mu^\pm} >5\gev$, $|y_{e^\pm}| <2.47$, $|y_{\mu^\pm}|<2.7$, $p_{T, \,l_1} >20\gev$, $p_{T, \,l_2} >20\gev$, \\
		& $\Delta R_{ll^\prime} >0.05$, $81\gev < M_{l^+ l^-} < 101\gev$, $M_{4l} >180\gev$  \\
		$ZZg$ & $ZZ$ selection, $p_{T, \,j} >20\gev$, $|y_{j}| <4.5$, $\Delta R_{lj} >0.4$\\
		$W^+ W^-$ &$p_{T, \,e^\pm} >27\gev$, $p_{T, \,\mu^\pm} >27\gev$, $|y_{l}| <2.5$, $p_{T, \,miss} >20\gev$, $M_{2l} >55\gev$,\\
		&  $M_{W^+W^-}>130\gev$, no jets with $p_{T,\,j}>35\gev$, $|\eta_j|<4.5$ \\
		$W^+W^-g$ & $W^+W^-$ lepton selection, one jet with $p_{T,j}>35\gev$, $|\eta_j|<4.5$, $\Delta R_{lj} >0.4$,\\
		& no further jets with  $p_{T,\,j}>35\gev$, $|\eta_j|<4.5$\\
		$Z\gamma$ & $p_{T, \,e^\pm} >25\gev$, $p_{T,\, \gamma} >30\gev$, $|y_{e^\pm}| <2.47$, $|y_{\gamma}|<2.37$, $p_{T, \,l_1} >30\gev$, $p_{T, \,l_2} >25\gev$, \\
		& $M_{ll}+M_{ll\gamma} > 182\gev$, $M_{l^+l^-} >40\gev$  \\				$Z\gamma g$ & $Z\gamma$ selection, $p_{T,\, j} >20\gev$, $|y_{j}| <4.5$, $\Delta R_{lj} >0.4$\\
		$Zh$ & $p_{T,\, \mu^\pm} >25\gev$, $|y_{\mu^\pm}|<2.7$, $p_{T,\,ll} >75\gev$, $81\gev < M_{l^+ l^-} < 101\gev$  \\
		& exactly two $b$-tagged jets with $p_{T, \,j} >20\gev$, $|y_{j}|<4.5$, $p_{T, \,j_1} >45\gev$, $\Delta R_{j_1 j_2} <\pi$, \\ 
		& $M_h>50\gev$\\
		$Zhg$ & $Zh$ selection, $p_{T, \,j} >20\gev$, $|y_{j}| <4.5$, $\Delta R_{lj} >0.4$\\
		\bottomrule
	\end{tabular}
	\caption{Phase space selections: if particles are numbered the particles are $p_T$ ordered and the respective selection criterium is only applied on the particle with the $n$th highest $p_T$. }
	\label{Tab.: Phase space selections}
\end{table}

Only events with exactly two ($W^+W^-$, $Zh$, $Z\gamma$) or four leptons ($ZZ$, $ZZg$) are taken into account. $Z\gamma$ events are forced to have at least one isolated photon, which is determined as described in Ref.~\cite{ATLAS:2022wnf}. 

Jets are reconstructed using the anti-$k_T$ algorithm~\cite{Cacciari:2008gp} with a radius parameter of $ R = 0.4$, if not stated otherwise. The Higgs boson in the $Zh$ process is obtained from exactly two b-tagged jets in the event. Perfect b-tagging is assumed. Beside the resolved setup defined in Tab.~\ref{Tab.: Phase space selections}, also a boosted, unresolved setup with $p_{T,Z} > 250~\mathrm{GeV}$ is investigated. In this setup, the Higgs boson is reconstructed from the large-$R$ jet $J$ with the highest $p_T$ fulfilling the following selection criteria: 
\begin{itemize}
	\item[-] radius parameter $R = 1.0 $,
	\item[-] among all small-$R$ jets $j$ with $p_{T,j} > 10~\mathrm{GeV}$,  $|y_j| <2.5$ and $R = 0.2$ within the large-$R$ jet, two of the three small-$R$ jets with the highest $p_T$ are $b$-tagged; all small-$R$ jets with $\Delta R_{jJ}<0.8$ are considered matched to the large-$R$ jet,
	\item[-] $p_{T,J} > 250~\mathrm{GeV}$, $m_J > 50~\mathrm{GeV}$.
\end{itemize}
If additional QED radiation is present, bare leptons are dressed via a cone dressing with resolution radius $R=0.1$.

Generated events are evaluated by making use of \Rivet4~\cite{Bierlich:2024vqo}.

\subsection{Integrated fiducial cross sections and polarisation fractions}
In this section, integrated fiducial cross sections for all investigated processes are presented. Both loop-induced inclusive and jet-associated processes are studied. Tables~\labelcref{Tab.: Fixed-order results two bosons,Tab.: Fixed-order results Zy,Tab.: Fixed-order results Zh} summarize the corresponding results at fixed LO. Two reference frames are considered for defining the polarisation states: the Lab and the COM frame. Tables~\labelcref{Tab.: Parton shower matched results ZZ,Tab.: Parton shower matched results WW,Tab.: Parton shower matched results Zy,Tab.: Parton shower matched results ZH,Tab.: Parton shower matched results ZH boosted} show the fiducial cross sections for all investigated loop-induced processes, matched either to a QCD PS alone or including additional QED radiation via YFS resummation. For brevity, these results are displayed for the COM frame only. 

\begin{table}[h]
	\centering
	\renewcommand{\arraystretch}{1.3}
	\begin{tabular}{lllllllll}
		\toprule
		& \multicolumn{2}{c}{$ZZ$} & \multicolumn{2}{c}{$ZZg$} & \multicolumn{2}{c}{$W^+W^-$} & \multicolumn{2}{c}{$W^+W^-g$} \\
		& XS~[fb] & \% & XS~[fb] & \% & XS~[fb] & \% & XS~[fb] & \%  \\
		\midrule
		full & 1.8323(5) & 97.8 & 1.1061(14) & 99.7 & $ 14.2372(28) $ & 98.6 &  & \\
		unpol &  1.8736(5) & 100  & 1.1089(17) & 100 & $ 14.442(4)$ & 100 &  5.233(12)                     &   100    \\
		\hline
		\multicolumn{9}{c}{Laboratory frame} \\
		\hline
		LL & 0.09966(8) & 5.32 & 0.06240(14) & 5.63 & $ 0.7709(5)$ & 5.34 &  0.3187(10)                   & 6.09 \\
		LT & 0.22356(10) & 11.9 & 0.14241(34)  & 12.8 & $ 1.9539(9 ) $ & 13.53 &  0.7349(29)                   & 14.04\\
		TL &  0.20843(9) & 11.1 & 0.13364(32) & 12.1 & $ 1.9562(9 ) $ & 13.55 &  0.7411(30)                   & 14.16 \\
		TT & 1.3226(4) &70.6 & 0.7650(12) & 69.0 & $ 10.3987(32) $ & 72.01  &  3.592(9)                     & 68.6\\
		int & 0.01935(8) & 1.03 & 0.00551(28) & 0.50 & -0.6381(15)& -4.42 & -0.153(6)                    & -2.93 \\
		\hline
		\multicolumn{9}{c}{Diboson centre-of-mass frame} \\
		\hline
		LL & 0.10117(8) & 5.40 & 0.06691(14) & 6.03 & $ 0.6183(9)$ & 4.28 & 0.3406(11)                   & 6.51\\
		LT &  0.039672(15) & 2.12 & 0.05046(17) & 4.55 & $ 0.43608   (17 ) $ & 3.01 & 0.3654(15)                   & 6.98\\
		TL &  0.039299(15) & 2.10 & 0.05118(10) & 4.47& $ 0.43310   (16 ) $ & 3.00 & 0.3654(15)                   & 6.98\\
		TT & 1.6949(5) &90.5 & 0.9445(15) & 85.2 & $ 13.326   (4 ) $ & 92.3 & 4.225(10)                    & 80.7 \\
		int & -0.001432(33) & -0.08 & -0.00239(16) & -0.22 & -0.3699(12) &-2.56 & -0.063(5)                    & -1.2 \\
		\bottomrule
	\end{tabular}
	\caption{Fiducial cross sections and polarisation fractions for several loop-induced diboson production processes at fixed LO. Polarised cross sections are calculated in the laboratory and in the diboson centre-of-mass frame.}
	\label{Tab.: Fixed-order results two bosons}
\end{table}

\begin{table}[t]
	\centering
	\renewcommand{\arraystretch}{1.3}
	\begin{tabular}{lllll}
		\toprule
		&  \multicolumn{2}{c}{$Z\gamma$} & \multicolumn{2}{c}{$Z\gamma g$} \\
		& XS~[fb] & \% & XS~[fb] & \% \\
		\midrule
		full & 7.7628(17) &99.8& 13.108(23) & 99.3\\
		unpol & $ 7.7789(17 ) $ & 100 & 13.203(21) & 100\\
		\hline
		\multicolumn{5}{c}{Laboratory frame} \\
		\hline
		L & $ 1.8730   (5 ) $ & 24.1 & 3.241(6)                     & 24.6 \\
		T & $ 5.9340   (14 ) $ & 76.3 & 10.045(17)                   & 76.1 \\
		int & -0.03011(15) & -0.39 &  -0.0832(20)                  & -0.63\\
		\hline
		\multicolumn{5}{c}{Diboson centre-of-mass frame} \\
		\hline
		L & $ 0.97530   (27 ) $ & 12.5 &1.8617(34)                   & 14.1\\
		T & $ 6.8750   (16 ) $ & 88.4 & 11.459(19)                   & 86.8\\
		int & -0.07341(9)& -0.94 & -0.1172(16)                  & -0.89\\
		\bottomrule
	\end{tabular}
	\caption{Like Tab.~\ref{Tab.: Fixed-order results two bosons} but for $Z\gamma$ production.}
	\label{Tab.: Fixed-order results Zy}
\end{table}

\begin{table}[t]
	\centering
	\renewcommand{\arraystretch}{1.3}
	\begin{tabular}{lllllll}
		\toprule
		&  \multicolumn{2}{c}{$Zh$} & \multicolumn{2}{c}{$Zh$ boosted} & \multicolumn{2}{c}{$Zhg$}\\
		& XS~[fb] & \% & XS~[fb] & \% & XS~[fb] & \% \\
		\midrule
		full & 0.80618(16) & 100 &0.028845(27) & 99.9 & 0.8135(13) & 100.2\\
		unpol & 0.80610(15) & 100 & 0.028869(27) & 100 & 0.8120(10) & 100\\
		\hline
		\multicolumn{7}{c}{Laboratory frame} \\
		\hline
		L & 0.68154(13) & 84.5 & 0.027636(26) & 95.7 & 0.6897(9)                    & 85.0\\
		T & 0.096613(33) & 12.0& 0.0010002(14)& 3.46 & 0.09768(23)                  & 12.0\\
		int & 0.027948(26)& 3.47& 0.0002323(32) & 0.80 & 0.02457(20)                  & 3.03\\
		\hline
		\multicolumn{7}{c}{Diboson centre-of-mass frame} \\
		\hline
		L & 0.80591(15) & 99.7 & 0.028555(6) & 98.9 & 0.8105(10)                   & 99.8\\
		T & 0.0010921(6) & 0.14& 0.0003501(7) & 1.21 & 0.003326(13)                 & 0.41\\
		int & -0.0008350(14) & -0.10 &-0.0008322(6) & -0.13 &-0.001828(26)                & -0.23\\
		\bottomrule
	\end{tabular}
	\caption{Like Tab.~\ref{Tab.: Fixed-order results two bosons} but for $Zh$ production.}
	\label{Tab.: Fixed-order results Zh}
\end{table}

\begin{table}[t]
	\centering
	\renewcommand{\arraystretch}{1.3}
	\begin{tabular}{lllllllll}
		\toprule
		& \multicolumn{4}{c}{$ZZ$} &\multicolumn{4}{c}{$ZZg$} \\
		& \multicolumn{2}{c}{QCD} & \multicolumn{2}{c}{QCD+QED} & \multicolumn{2}{c}{QCD} & \multicolumn{2}{c}{QCD+QED} \\
		& XS~[fb] & \% & XS~[fb] & \% & XS~[fb] & \%  & XS~[fb] & \% \\
		\midrule
		full & 1.8530(5) & 97.7  & 1.7353(8) & 98.0 & 2.6154(15) & 99.0 & 2.4534(21)& 98.7 \\
		unpol & 1.8966(4) & 100  & 1.7707(36) & 100  & 2.643(5) & 100& 2.486(6)  & 100  \\
		\hline
		LL & 0.10225(14)                  & 5.39 &  0.09488(7)& 5.36 &   0.1488(4)                    & 5.63 & 0.1404(9)                    & 5.65\\
		LT &  0.040117(13) & 2.12 & 0.037424(10) & 2.11  & 0.08932(25)                  & 3.38 &0.08362(25)                  & 3.36 \\
		TL &  0.039764(11) & 2.10 & 0.037088(10) & 2.09  & 0.08789(25)                  & 3.33  & 0.08263(25)                  & 3.32\\
		TT & 1.71569(33) &90.5 & 1.60238(32) & 90.5  &2.320(5)                     & 87.8  &2.182(5)                     & 87.8\\
		int & -0.001206(24)                & -0.06 &-0.001050(23)& -0.06 & -0.0028(4)                   & -0.11 &-0.0030(4)                   & -0.12 \\
		\bottomrule
	\end{tabular}
	\caption{Fiducial cross sections and polarisation fractions for loop-induced, inclusive and jet-associated $ZZ$-production at LO matched to the parton shower. Polarised cross sections are calculated in the diboson centre-of-mass frame.}
	\label{Tab.: Parton shower matched results ZZ}
\end{table}
\begin{table}[t]
	\centering
	\renewcommand{\arraystretch}{1.3}
	\begin{tabular}{lllllllll}
		\toprule
		& \multicolumn{4}{c}{$W^+W^-$} &\multicolumn{4}{c}{$W^+W^-g$} \\
		& \multicolumn{2}{c}{QCD} & \multicolumn{2}{c}{QCD+QED} & \multicolumn{2}{c}{QCD} & \multicolumn{2}{c}{QCD+QED}  \\
		& XS~[fb] & \% & XS~[fb] & \% & XS~[fb] & \%  & XS~[fb] & \%  \\
		\midrule
		full & 10.264(5) & 98.4 & 10.044(5) & 98.3 &&&&\\
		unpol & 10.4173(20) & 100 & 10.2144(25)& 100 & 8.368(20) & 100 & 8.269(20) & 100\\
		\hline
		LL & 0.4594(4) & 4.41 &0.4436(5)                    & 4.34 & 0.4324(19) & 5.17 & 0.4233(19) & 5.12\\
		LT & 0.31532(12) & 3.03 & 0.30877(10)                  & 3.02 & 0.4008(17) & 4.79 & 0.3965(18) & 4.80\\
		TL &  0.31458(12) & 3.02 & 0.30825(10)                  & 3.02 & 0.3998(17) & 4.78 & 0.3921(16) &4.74\\
		TT & 9.6429(27) & 92.6 & 9.4614(23) & 92.6 & 7.295(18) & 87.2 & 7.207(19) & 87.2\\
		int & -0.3150(8) & -3.02 & -0.3077(7)                   & -3.01 & -0.160(7) & -1.91 & -0.150(7) & -1.81\\
		\bottomrule
	\end{tabular}
	\caption{Like Tab.~\ref{Tab.: Parton shower matched results ZZ} but for $W^+W^-$ production.\\
		For off-shell jet-associated $\PW^+ \PW^-$ production we have been unable to receive a stable integration result which is why we do not show a result for this process.}
	\label{Tab.: Parton shower matched results WW}
\end{table}

\begin{table}[t]
	\centering
	\renewcommand{\arraystretch}{1.3}
	\begin{tabular}{lllllllll}
		\toprule
		& \multicolumn{4}{c}{$Z\gamma$} & \multicolumn{4}{c}{$Z\gamma g$}   \\
		& \multicolumn{2}{c}{QCD} & \multicolumn{2}{c}{QCD+QED} & \multicolumn{2}{c}{QCD} & \multicolumn{2}{c}{QCD+QED} \\
		& XS~[fb] & \% & XS~[fb] & \% & XS~[fb] & \% & XS~[fb] & \% \\
		\midrule
		full & 9.3778(34) & 108.5 & 9.1882(33) & 108.9 & 14.67(4) & 108.9 &  14.355(40)& 108.9\\
		unpol &  $ 8.6455(23 ) $ & 100 & 8.4398(20) & 100 & 13.467(26) & 100 & 13.186(26) & 100\\
		\hline
		L & $ 1.05931   (35 ) $ & 12.3 & 1.03388(30) &12.3 &1.868(4)                     & 13.9&1.833(5)                     & 13.9\\
		T & $ 7.6798   (20 ) $ & 88.8 & 7.4979(17) & 88.8 &11.720(23)                   & 87.0 &11.475(23)                   & 87.0\\
		int & -0.09364(13)& -1.08 & -0.09196(11) & -1.09 &-0.1212(18)                  & -0.90&-0.1218(18)                  & -0.92\\
		\bottomrule
	\end{tabular}
	\caption{Like Tab.~\ref{Tab.: Parton shower matched results ZZ} but for $Z\gamma$ production.}
	\label{Tab.: Parton shower matched results Zy}
\end{table}

\begin{table}[t]
	\centering
	\renewcommand{\arraystretch}{1.3}
	\begin{tabular}{lllllllll}
		\toprule
		& \multicolumn{4}{c}{$Zh$} & \multicolumn{4}{c}{$Zhg$}  \\
		& \multicolumn{2}{c}{QCD} & \multicolumn{2}{c}{QCD+QED} & \multicolumn{2}{c}{QCD} & \multicolumn{2}{c}{QCD+QED}   \\
		& XS~[fb] & \% & XS~[fb] & \% & XS~[fb] & \%  & XS~[fb] & \%  \\
		\midrule
		full & 0.68813(16) & 100.3& 0.68813(16)& 103.2& 1.4554(15) & 100 & 1.4554(15)& 102.9\\
		unpol & 0.68596(12) & 100 & 0.66651(10)&100 & 1.4553(18)  & 100 & 1.4139(17) & 100\\
		\hline
		L & 0.68569(12) & 100 & 0.66624(10) & 100 & 1.4536(18)                   & 99.9& 1.4122(17)                   & 99.9\\
		T & 0.0009755(5) & 0.14 & 0.0009508(4)& 0.14 & 0.004150(17)                 & 0.29& 0.004057(17)                 & 0.29\\
		int & -0.0006968(12) & 0.10 & -0.0006792(10)& 0.10 & -0.002400(35)                & -0.16 & -0.00231(4)                  & -0.16\\
		\bottomrule
	\end{tabular}
	\caption{Like Tab.~\ref{Tab.: Parton shower matched results ZZ} but for $Zh$ production.}
	\label{Tab.: Parton shower matched results ZH}
\end{table}

\begin{table}[t]
	\centering
	\renewcommand{\arraystretch}{1.3}
	\begin{tabular}{lllll}
		\toprule
		& \multicolumn{4}{c}{$Zh$ boosted}  \\
		& \multicolumn{2}{c}{QCD} & \multicolumn{2}{c}{QCD+QED} \\
		& XS~[fb] & \% & XS~[fb] & \% \\
		\midrule
		full & 0.025452(22)& 100.1 & 0.025452(28)& 102.7\\
		unpol & 0.025418(12) & 100& 0.024790(19)& 100\\
		\hline
		L & 0.025122(22) &  98.8 & 0.024500(19)& 98.8\\
		T & 0.00032811(38)& 1.29 & 0.00032065(33)& 1.29\\
		int & -0.0000318(6)& -0.12 & -0.0000311(5)& -0.14\\
		\bottomrule
	\end{tabular}
	\caption{Like Tab.~\ref{Tab.: Parton shower matched results ZZ} but for boosted $Zh$ production.}
	\label{Tab.: Parton shower matched results ZH boosted}
\end{table}

\subsubsection{Inclusive loop-induced production }
At fixed order, the on-shell calculations closely reproduce the off-shell predictions for inclusive loop-induced processes, with differences compatible with the expected intrinsic NWA uncertainty of $\mathcal{O}(\Gamma/m)$.
Consistent with previous studies~\cite{Denner:2020eck,Denner:2020bcz,Denner:2021csi,Denner:2022riz,Denner:2023ehn, Dao:2023pkl, Dao:2023kwc,Dao:2024ffg, Le:2022lrp, Le:2022ppa, Poncelet:2021jmj, Carrivale:2025mjy, Pelliccioli:2025com}, the pure transverse contribution dominates in diboson production processes without an intermediate Higgs boson with about 70-76\% in the Lab frame definition and 88-92\% in the COM. In contrast, the longitudinal component dominates in $\PZ\PH$ production, ranging between 80\% to nearly 100\% depending on the setup. Compared to their quark-induced counterparts at \nnloqcd~\cite{Carrivale:2025mjy, Poncelet:2021jmj}, the loop-induced $\PZ \PZ$ and the $\PW^+ \PW^-$ production processes feature a significantly larger TT contribution (72\% vs. 50\%) compensated by smaller mixed contributions. 

Different polarisation definitions affect the individual polarisation contributions differently depending on the process. In processes without an intermediate Higgs boson, the longitudinal fraction is reduced in the COM frame relative to the Lab frame. In loop-induced $\PZ\PZ$ and $\PW^+\PW^-$ production, the mixed-polarisation contribution is particularly reduced. However, for loop-induced $\PZ\PH$ production, the $\PZ$ boson is almost entirely longitudinally polarised in the COM frame, compared to $\sim85\%$ in the Lab frame. In the boosted setup for $\PZ\PH$ production, the two polarisation definitions yield more similar results as the two bosons become more nearly back-to-back than in the resolved setup. The longitudinal polarisation contribution significantly increases in the Lab frame and decreases slightly in the COM definition compared to the resolved setup. 

Interference contributions are generally small, at the level of $1\%$ or less, and are usually smaller in the COM frame than in the Lab frame (except for $\PZ\gamma$ production). A sizeable interference contribution is only found for loop-induced $\PW^+\PW^-$ production, where the interference contribution is comparable in magnitude to the pure longitudinal contribution ($-2.5\%$ to $-4.4\%$). This behaviour may originate from the tighter lepton-$p_T$ selection compared to the $\PZ\PZ$ process. 

Including resummation effects significantly increases the difference between on- and off-shell calculations for $\PZ\gamma$ and $\PZ\PH$ production from sub-percent differences to deviations of $9\%$ and $3\%$, respectively. Additional QED radiation accounts for most of the difference in $\PZ\PH$ production, while the largest deviation in $\PZ\gamma$ production is caused by QCD resummation effects.

When comparing fixed-order and PS-matched results, the cross sections change by $2-6\%$ ($\PZ\PZ$), $10\%$ ($\PZ\gamma$), and $21\%$ ($\PZ\PH$) due to resummation effects. The dominant effect comes from QCD resummation for all but $\PZ \PZ$ production and is mostly negative. Only in $\PZ\gamma$ production does the cross section increase. For $\PW^+\PW^-$ production, the difference between the fixed-order and resummed results is particularly large ($\sim -30 \%$), which is mostly due to the strict jet selection. Additional QED radiation tends to decrease the difference to the fixed-order result for $\PZ\PZ$ and $\PZ\gamma$ production, while for $\PZ\PH$ and $\PW^+\PW^-$ production it further reduces the cross section compared to fixed order.

As was previously observed in the case of quark-induced production of two massive gauge bosons~\cite{Pelliccioli:2023zpd}, the polarisation decomposition remains largely unaffected by resummation effects. The polarisation fractions are almost unchanged, deviations from the fixed-order fractions remain below the percent level. 

\subsubsection{Jet-associated loop-induced production }
The fixed-order cross sections of jet-associated loop-induced processes are sizeable relative to those of the corresponding inclusive processes, amounting to approximately $60\%$ ($\PZ\PZ$), $36\%$ ($\PW^+\PW^-$), $100\%$ ($\PZ\PH$), and even $170\%$ for $\PZ\gamma$ production. Differences in the polarisation decomposition are small in the Lab frame definition, with differences below one percentage point. The only exception is $\PW^+\PW^-\Pg$, where the pure transverse polarisation component and the interference contribution show larger differences between the inclusive and jet-associated processes with a difference in percentages points of -3.5 and 1.5\%. For the COM polarisation definition, significantly larger differences in the polarisation fractions are found between the two process types, especially in processes involving two massive gauge bosons, where the mixed polarisation contribution more than doubles. This indicates that additional hard QCD radiation can substantially modify the polarisation composition and motivates the inclusion of higher-order corrections in future precision predictions. Regardless of the frame definition, a larger longitudinal contribution is found in the jet-associated processes compared to their inclusive counterparts. 

In contrast to the inclusive processes, incorporating resummation effects substantially increases the fiducial cross sections by between 60-140~\% compared to the fixed-order result for all the examined processes. The only exception is $\PZ\gamma$ production, where the enhancement is minimal. The agreement between on- and off-shell calculations is similar to that observed for the PS-matched inclusive results. The polarisation fractions for processes involving two massive intermediate vector bosons also exhibit significant changes due to resummation effects, favouring transverse over longitudinal polarisation. Mixed fractions are influenced most, showing a reduction of about 30\%. The dominant effect on both the polarisation fractions and the unpolarised cross section, compared to fixed order, arises from the inclusion of additional QCD radiation. Like for the unpolarised cross section, polarisation fractions of $\PZ\gamma$ production are only mildly influenced by QCD resummation. QED radiation reduces the overall cross section by at most $4\%$, and has a minor effect on the polarisation decomposition, with differences amounting to less than $1\%$.

\subsection[Differential results]{Differential results}
In the remainder of this section, we exemplify the effect on differential distributions for the processes under investigation. We have selected polarisation-sensitive observables for which the effect of different polarisation definitions or resummation contributions is significant. Frame-dependent effects are studied at fixed LO. For the discussion of resummation effects, we consider polarised cross sections defined only in the COM frame. QCD and QED resummation effects are shown separately. For brevity, we only display a few observables per process, so the selection is not exhaustive. 

\subsection{Loop-induced inclusive $\PZ\PZ$ production}
\begin{figure}[!t]
	\centering
	\begin{subfigure}{0.465\textwidth}
		\begin{subfigure}{\textwidth}
			\includegraphics[width=\textwidth]{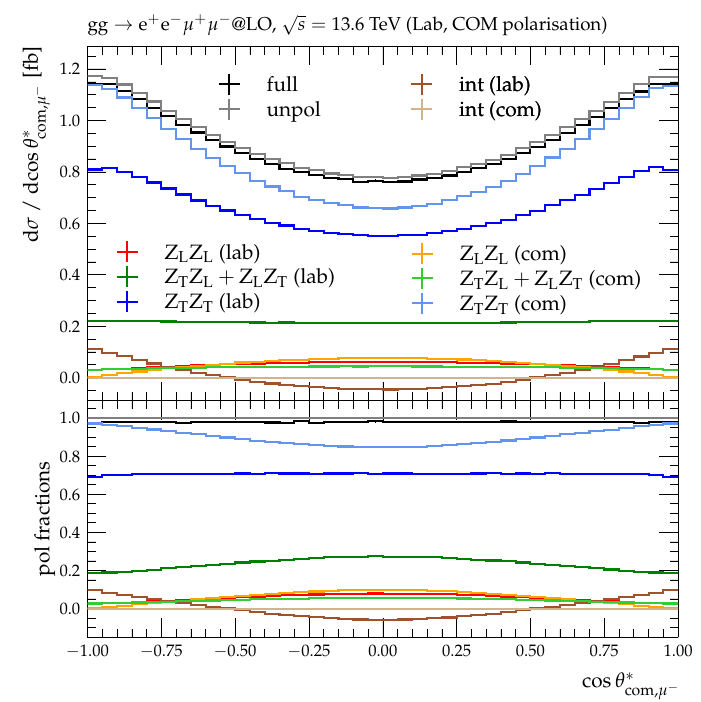}
		\end{subfigure}
	\end{subfigure}
	\begin{subfigure}{0.465\textwidth}
		\begin{subfigure}{\textwidth}
			\includegraphics[width=\textwidth]{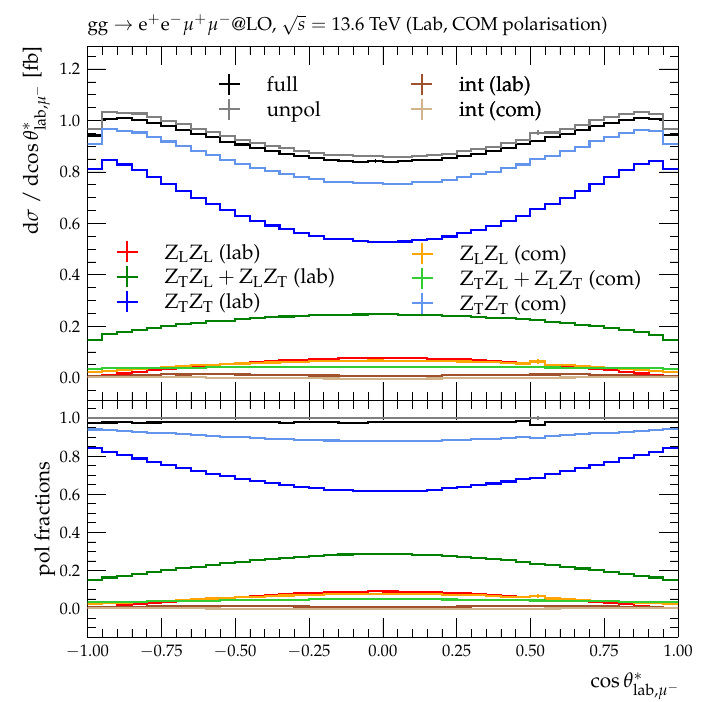}
		\end{subfigure}
	\end{subfigure}
	\hspace*{0.05\textwidth}
	\begin{subfigure}{0.465\textwidth}
		\begin{subfigure}{\textwidth}
			\includegraphics[width=\textwidth]{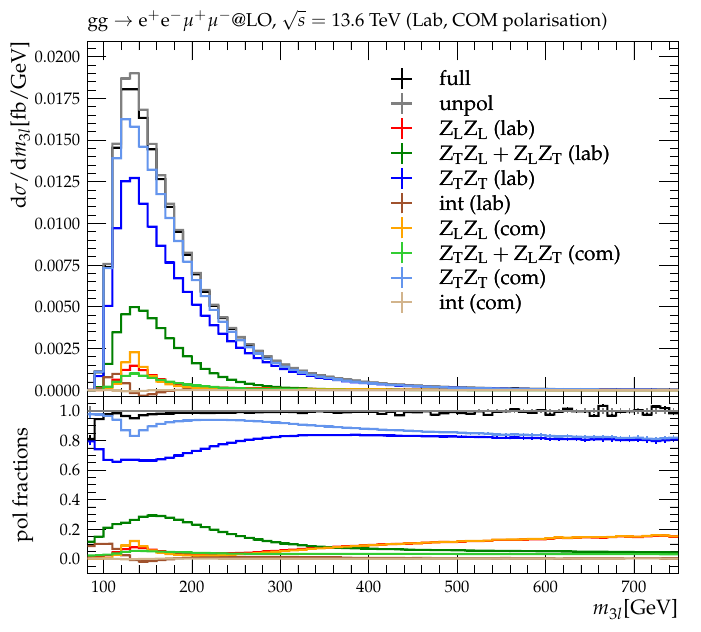}
		\end{subfigure}
	\end{subfigure}
	\caption{Polarised differential cross sections of the muon decay angle defined in the diboson centre-of-mass frame~$\cos \theta^*_{\mathrm{com},\mu^-}$~(top left) and in the laboratory frame~~$\cos \theta^*_{\mathrm{lab},\mu^-}$~(top right), as well as the invariant mass of the electron, positron and muon system $m_{3l}$~(bottom) for loop-induced, inclusive $\PZ\PZ$ boson production at LO. Polarisation is calculated in the laboratory~(lab) and in the diboson centre-of-mass (com) frame. Upper panels display the absolute distributions, lower panels show the differential polarisation fractions with respect to the unpolarised result.}
	\label{Fig.: differential XS ZZ FO}
\end{figure}

We start with inclusive $\PZ \PZ$ production. For this process, fixed-LO distributions for the muon decay angle $\cos \theta^*_{\mu^-}$, defined based on the vector boson direction in the Lab and in the COM frame, and the invariant mass of the electron-positron pair and the muon $m_{3l}$ are displayed in Fig.~\ref{Fig.: differential XS ZZ FO}. The difference between the on- and off-shell calculations is also small at the differential level. It is entirely flat for $\cos \theta^*_{\mu^-}$, whereas for $m_{3l}$ small, non-trivial differences in the peak region are found. The on-shell result leads to a higher peak cross section and a more gradual rise below the peak.

The pure transverse polarisation contribution dominates over the entire phase space for all three observables. The lepton decay angle is one of the most polarisation-sensitive observables, with very different shapes for the pure transverse polarisation contribution, which has a maximum near $\cos \theta^*_{\mu^-}=1$, and for the longitudinal contributions which peak at $\cos \theta^*_{\mu^-}=0$. The distributions of the pure longitudinal state in both polarisation definitions, and of the mixed polarisation contribution in the COM frame, exhibit a very similar behaviour for both angle definitions. The pure transverse polarisation contribution has a deeper minimum around $\cos \theta^*_{\mu^-}=0$ and higher maxima if the frame choice for the observable and for the polarisation definition coincide. For differing choices, its shape is similar to or identical to the unpolarised distribution. The interference (mixed) contribution, defined in the Lab frame, is entirely flat when the frame choices (dis)agree.    

For $m_{3l}$, the differences in shape between the various polarisations are smaller. Nonetheless, the transverse and longitudinal fractions peak at different values of $m_{3l}$.The polarisation definition has a pronounced effect on this distribution. While pure longitudinal polarisation contributions differ only slightly in the peak region between the Lab and COM definitions, pure transverse and mixed polarisation states show substantial differences in this kinematic region. In the Lab frame definition, the pure transverse polarisation contribution is significantly reduced at the peak, reaching only around $60\%$, compensated by a high mixed polarisation fraction of around $30\%$. For high invariant masses, the two definitions of polarisation approach each other. The pure longitudinal polarisation contribution increases in this region, while the mixed (pure transverse) contribution is reduced in the Lab (COM) frame definition. Only in the Lab-frame definition does the interference exhibit a non-trivial shape.

\begin{figure}[!t]
	\centering
	\begin{subfigure}{0.465\textwidth}
		\begin{subfigure}{\textwidth}
			\includegraphics[width=\textwidth]{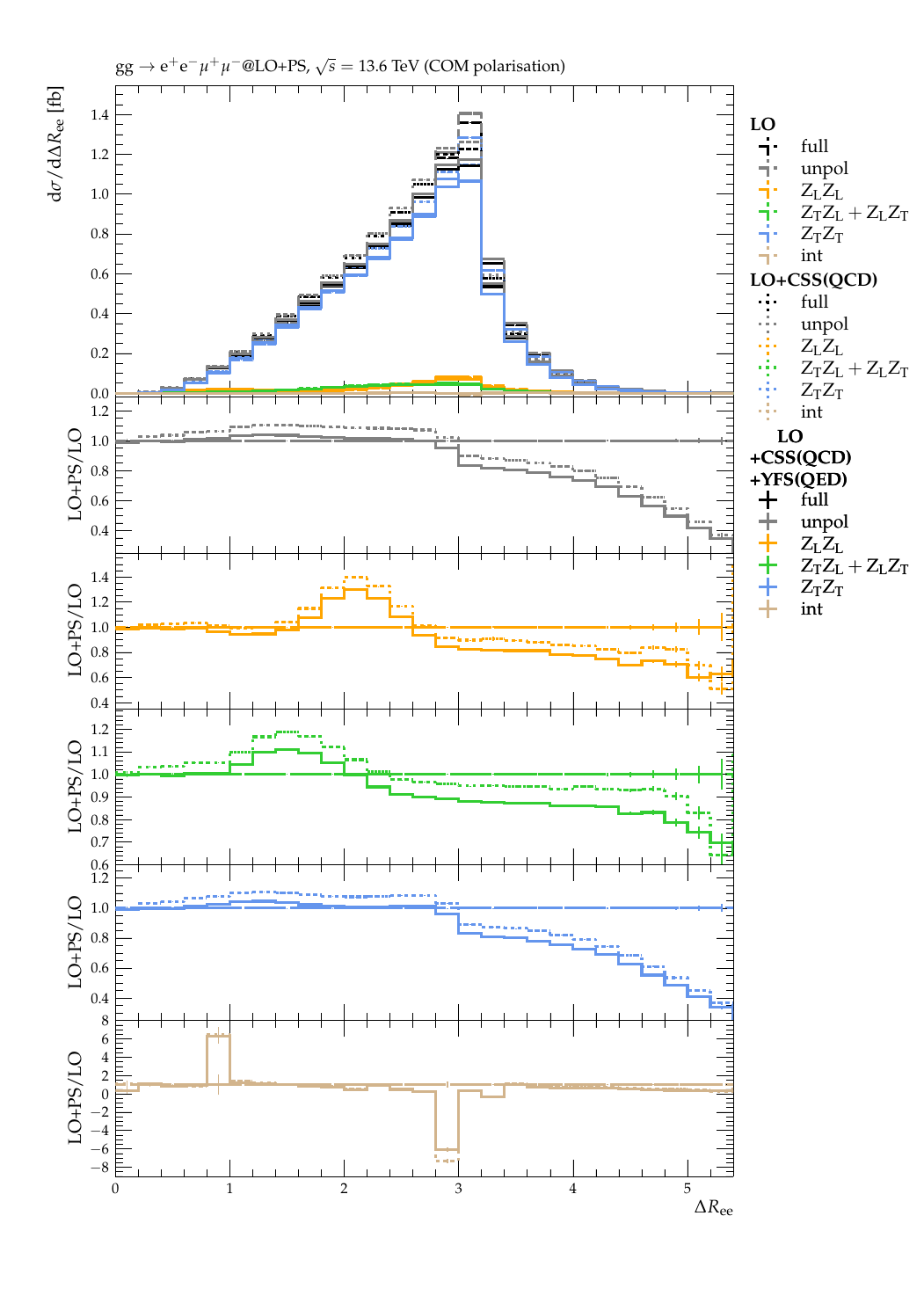}
		\end{subfigure}
	\end{subfigure}
	\hspace*{0.025\textwidth}
	\begin{subfigure}{0.465\textwidth}
		\begin{subfigure}{\textwidth}
			\includegraphics[width=\textwidth]{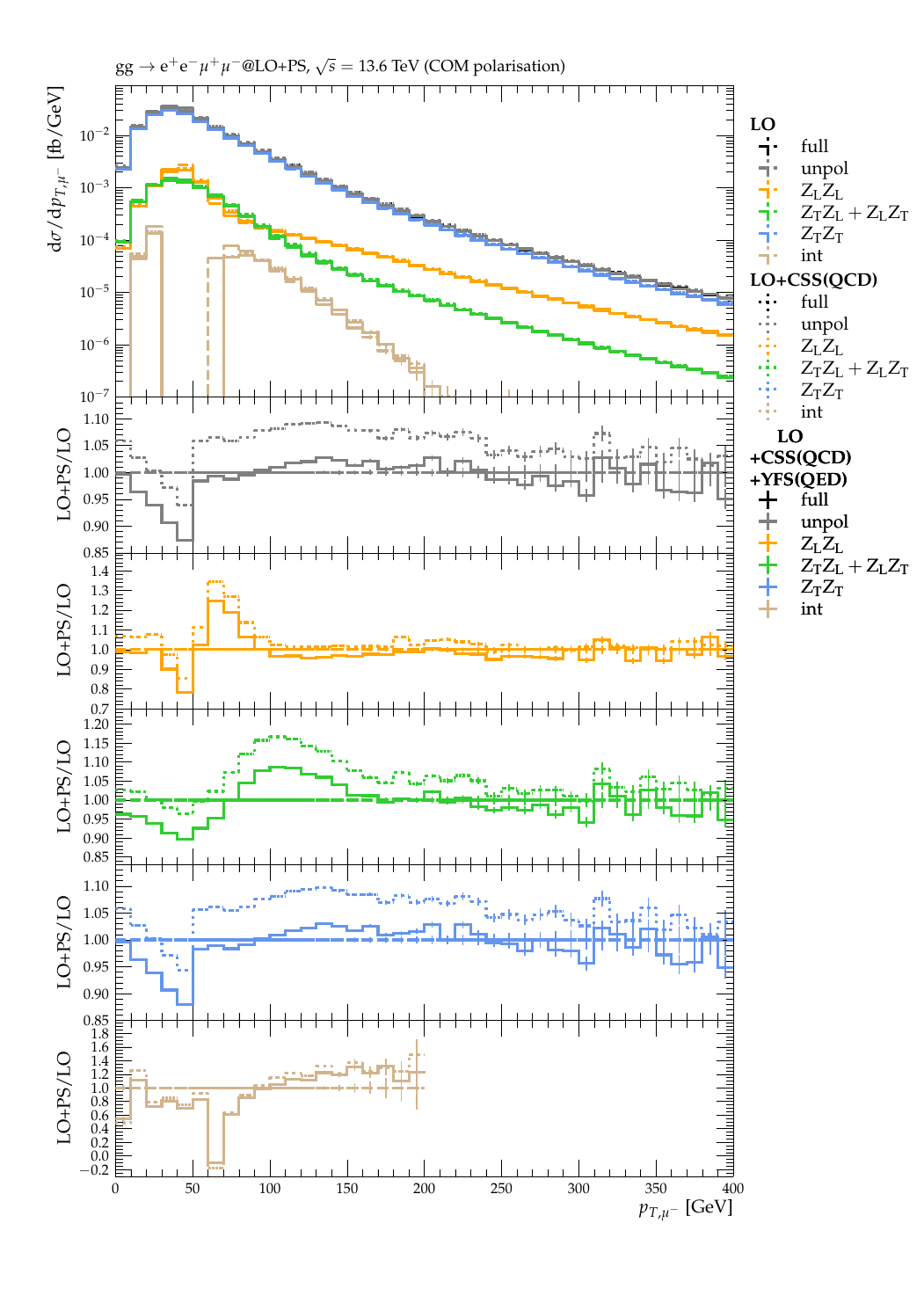}
		\end{subfigure}
	\end{subfigure}
	\caption{Polarised differential cross sections of the angular separation~$\Delta R_{\Pe\Pe}$(left) between the electron and positron, and the transverse momentum of the muon $p_{T, \mu^-}$~(right) for loop-induced, inclusive $\PZ\PZ$ boson production at LO, LO matched to a QCD shower only (LO+CSS(QCD)) and LO matched to QCD and QED resummation (LO+CSS(QCD)+YFS(QED)). Polarisation is calculated in the diboson centre-of-mass~(com) frame. Upper panels display the absolute distributions, lower panels show the cross sections including PS effects with respect to their fixed-order counterparts for the on-shell unpolarised calculation and the different polarisation contributions. If not shown, interference contributions are zero in that region.} 
	\label{Fig.: differential XS ZZ LOPS}
\end{figure}

The angular separation $\Delta R_{\Pe\Pe}$ between electron and positron and the transverse momentum of the muon $p_{T,\mu^-}$ are shown in Fig.~\ref{Fig.: differential XS ZZ LOPS} with and without PS effects. The ratio panels compare the \lo and \lops cross sections for different polarisation contributions and for the unpolarised result. The influence of resummation effects on the polarisation fractions can thus be estimated by comparing the shape of the ratios for polarised and unpolarised distributions. The largest shape differences compared to the fixed-order result are caused by QCD radiation and reach up to 60\% and 15\% at large $\Delta R_{\Pe\Pe}$ and small $p_{T,\mu^-}$, respectively. In contrast to the integrated polarisation fractions, some differential polarisation fractions change substantially once QCD shower effects are included. While the pure transverse polarisation fraction remains almost unchanged, the pure longitudinal (mixed) contribution is significantly enhanced around $\Delta R_{ee}=2.1(1.5)$ and $p_{T,\mu^-}=70~\gev$ $(120~\gev)$ as well as for large angular separations. The interference contribution shows large changes at small $p_{T,\mu^-}$ values and around $\Delta R_{\Pe\Pe}=2.9$. Additional QED radiation mainly produces an overall cross section reduction that is similar for all polarisation components in agreement with the integrated result. 

\subsection{Differential results for loop-induced jet-associated $ZZ$ production}
\begin{figure}[t]
	\centering
	\begin{subfigure}{0.465\textwidth}
		\begin{subfigure}{\textwidth}
			\includegraphics[width=\textwidth]{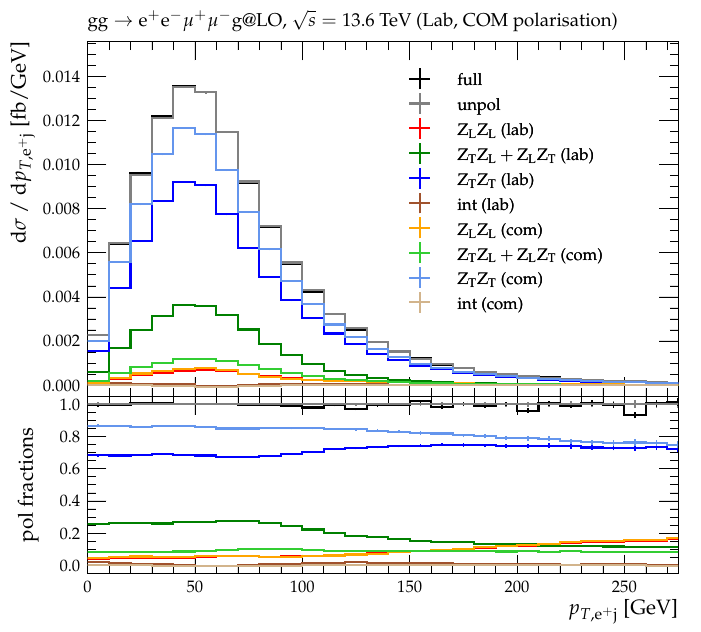}
		\end{subfigure}
	\end{subfigure}
	\begin{subfigure}{0.465\textwidth}
		\begin{subfigure}{\textwidth}
			\includegraphics[width=\textwidth, trim=0.6cm 1.5cm 0cm 1.1cm, clip]{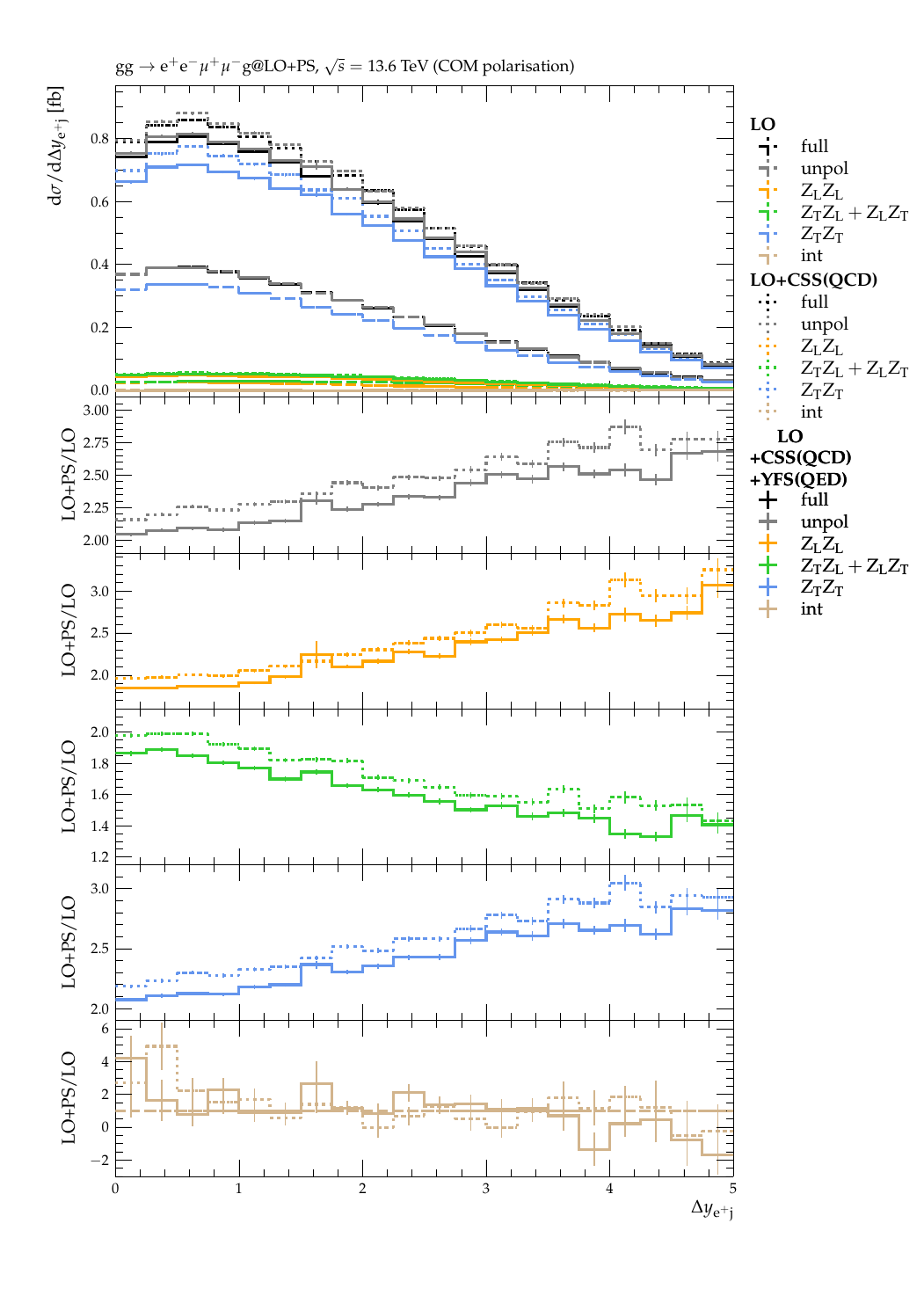}
		\end{subfigure}
	\end{subfigure}
	\caption{Polarised differential cross sections of the transverse momentum of, and the rapidity difference between, the jet and the positron, $p_{T,\Pe^+\Pj}$~(left) and $\Delta y_{\Pe^+\Pj}$~(right), for loop-induced $\PZ\PZ\Pg$ boson production at LO, LO matched to a QCD shower only (LO+CSS(QCD)) and LO matched to QCD and QED resummation (LO+CSS(QCD)+YFS(QED)). Polarisation is calculated in the laboratory~(lab) and in the diboson centre-of-mass~(com) frame. Upper panels display the absolute distributions, the lower-left panel shows the differential polarisation fractions with respect to the unpolarised result, the lower-right panels show the cross sections including PS effects with respect to their fixed-order counterparts for the on-shell unpolarised calculation and the different polarisation contributions.}
	\label{Fig.: differential XS ZZj FO}
\end{figure}

Fig.~\ref{Fig.: differential XS ZZj FO} presents two observables for loop-induced $\PZ\PZ$ production that can only be defined for the jet-associated process. The transverse momentum of the positron and the jet $p_{T,{\mathrm{e}^+\mathrm{j}}}$ is shown at fixed order for two different polarisation definitions. The rapidity separation between the positron and the hardest jet $\Delta y_{\mathrm{e}^+\mathrm{j}}$ is used to illustrate the PS effects on this process.

The on-shell calculation agrees very well with the off-shell result. The dependence on the polarisation definition is similar to that observed for $m_{3l}$ in the inclusive case. In the bulk of the distribution, the pure transverse contribution is reduced in the Lab compared to the COM definition, featuring a significantly increased mixed-polarisation fraction in this region. The contribution of the pure longitudinal state and the interference coincide for both polarisation definitions. TT and TL+LT contributions defined in the different frames approach each other in the high-$p_T$ tails of the distribution. 

Resummation effects, particularly those from QCD radiation, significantly affect the polarisation fractions over large ranges of phase space. The cross sections of the pure polarisation states are affected in qualitatively similar ways by the resummation effects, producing a roughly linear increase of the resummed cross section with increasing rapidity separation. The slopes differ, however, leading to modified polarisation fractions compared to fixed order. The mixed polarisation states and the interference contribution behave entirely differently. The interference contribution is essentially unchanged relative to fixed order, while the TL+LT cross section decreases with increasing rapidity separation.

\begin{figure}
	\centering
	\begin{subfigure}{0.465\textwidth}
		\begin{subfigure}{\textwidth}
			\includegraphics[width=\textwidth]{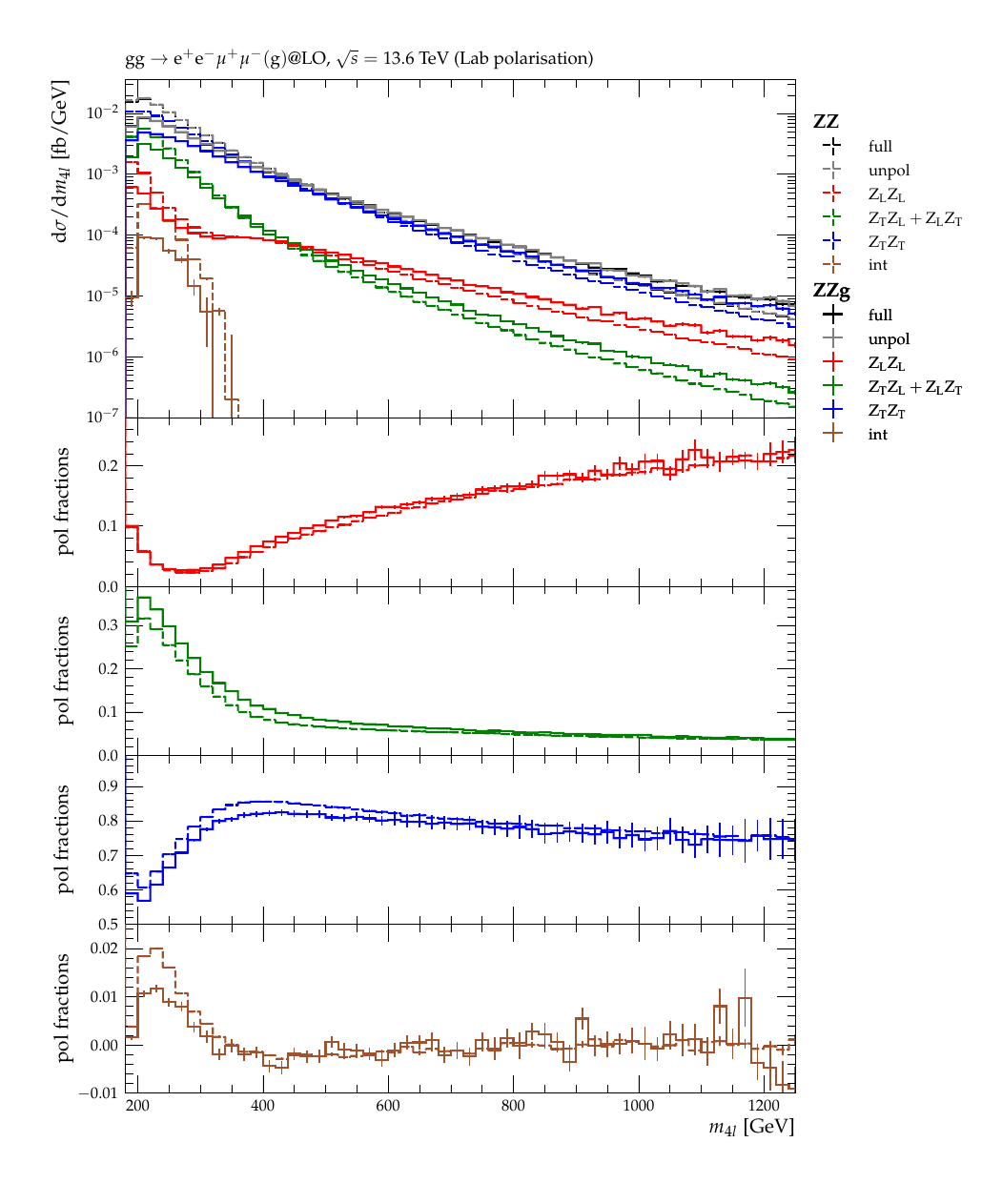}
		\end{subfigure}
	\end{subfigure}
	\hspace*{0.025\textwidth}
	\begin{subfigure}{0.465\textwidth}
		\begin{subfigure}{\textwidth}
			\includegraphics[width=\textwidth]{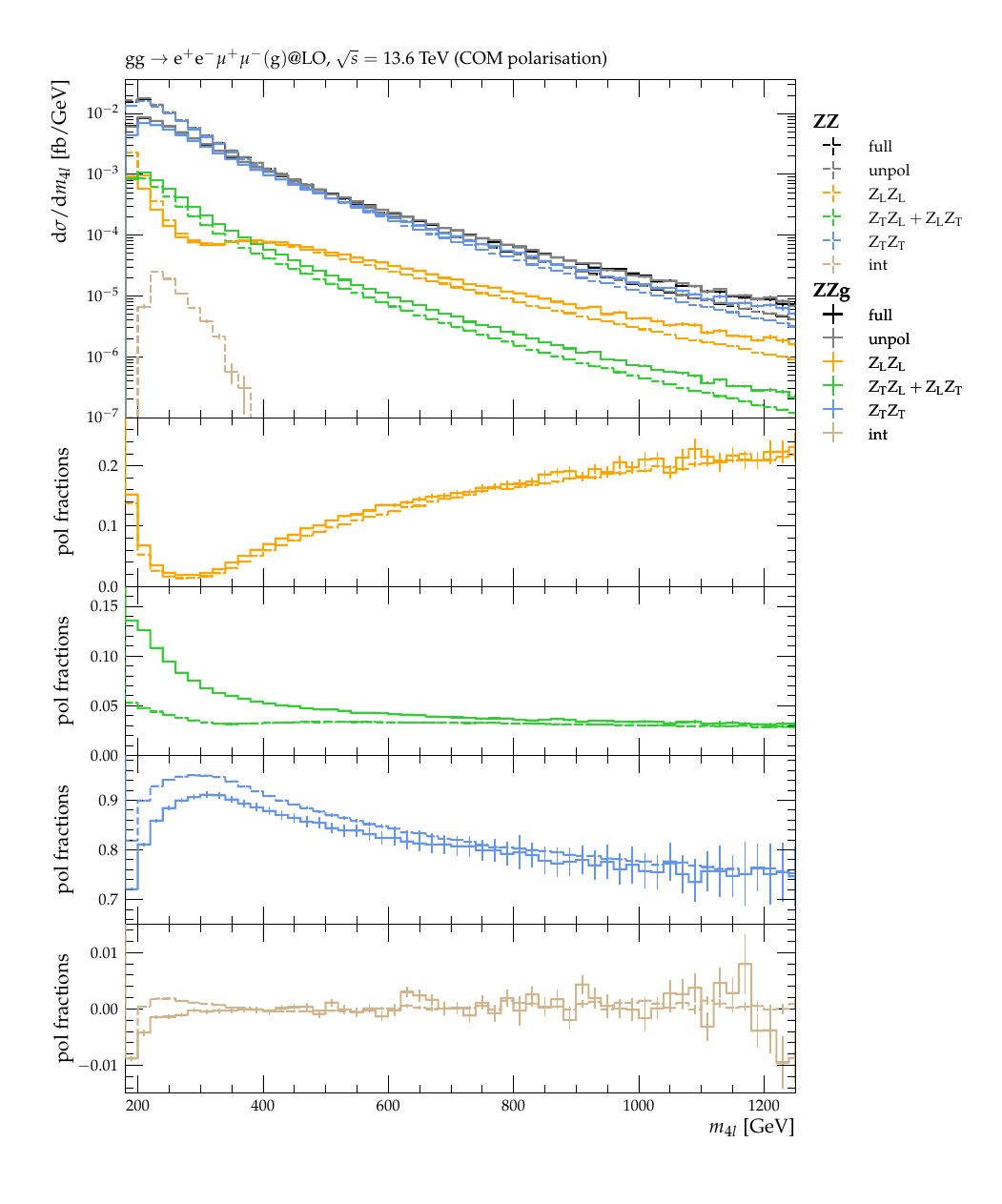}
		\end{subfigure}
	\end{subfigure}
	\caption{Polarised differential cross sections of the $ZZ$ invariant mass $m_{4l}$ for loop-induced, inclusive and jet-associated $ZZ$ boson production at LO. Polarisation is calculated in the laboratory frame~(lab, left) and in the diboson centre-of-mass frame~(com, right). Upper panels display the absolute distributions, lower panels show the differential polarisation fractions with respect to the unpolarised result of the respective process.}
	\label{Fig.: differential XS ZZ ZZj mass}
\end{figure}

\begin{figure}
	\centering
	\begin{subfigure}{0.465\textwidth}
		\begin{subfigure}{\textwidth}
			\includegraphics[width=\textwidth]{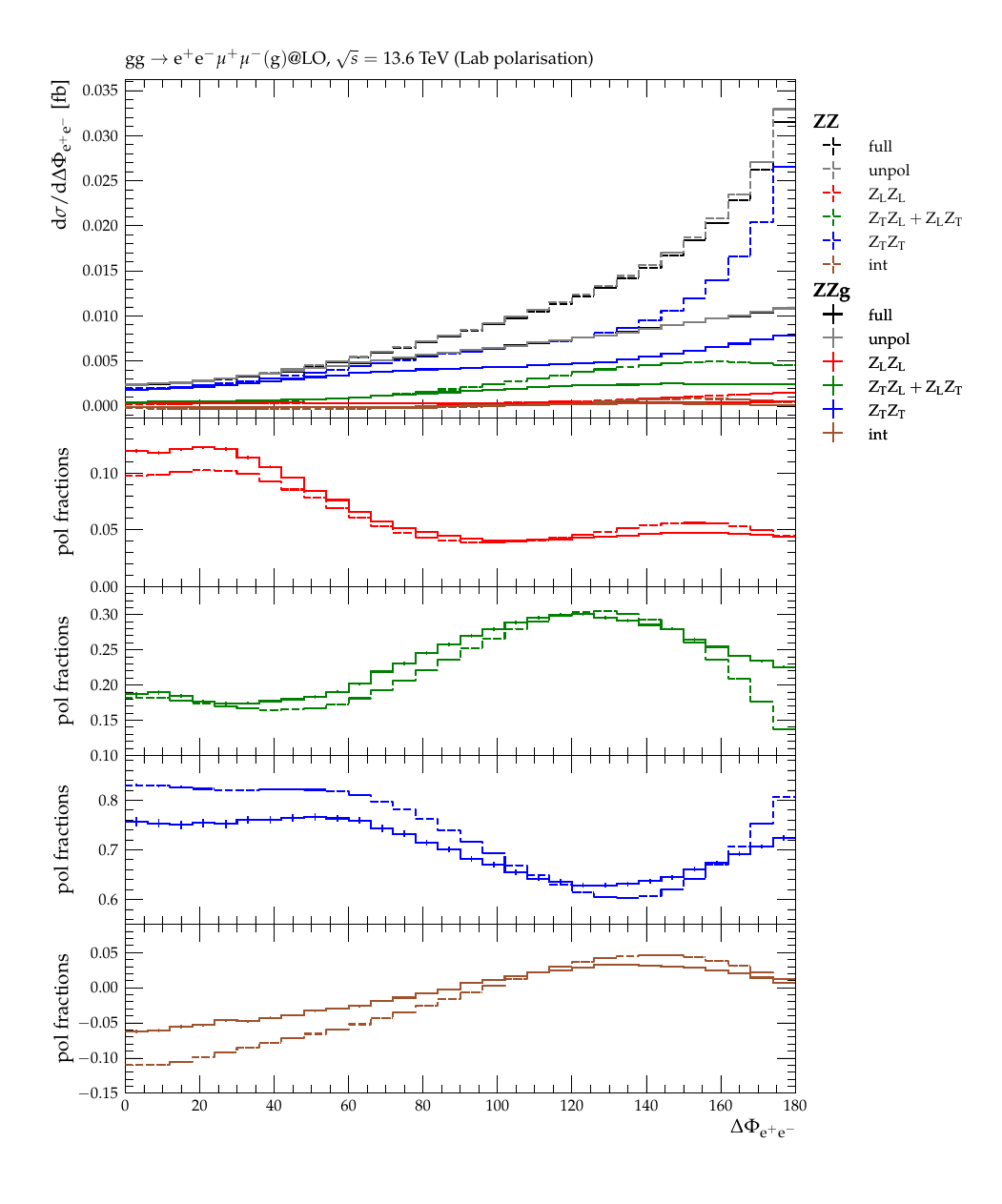}
		\end{subfigure}
	\end{subfigure}
	\hspace*{0.025\textwidth}
	\begin{subfigure}{0.465\textwidth}
		\begin{subfigure}{\textwidth}
			\includegraphics[width=\textwidth]{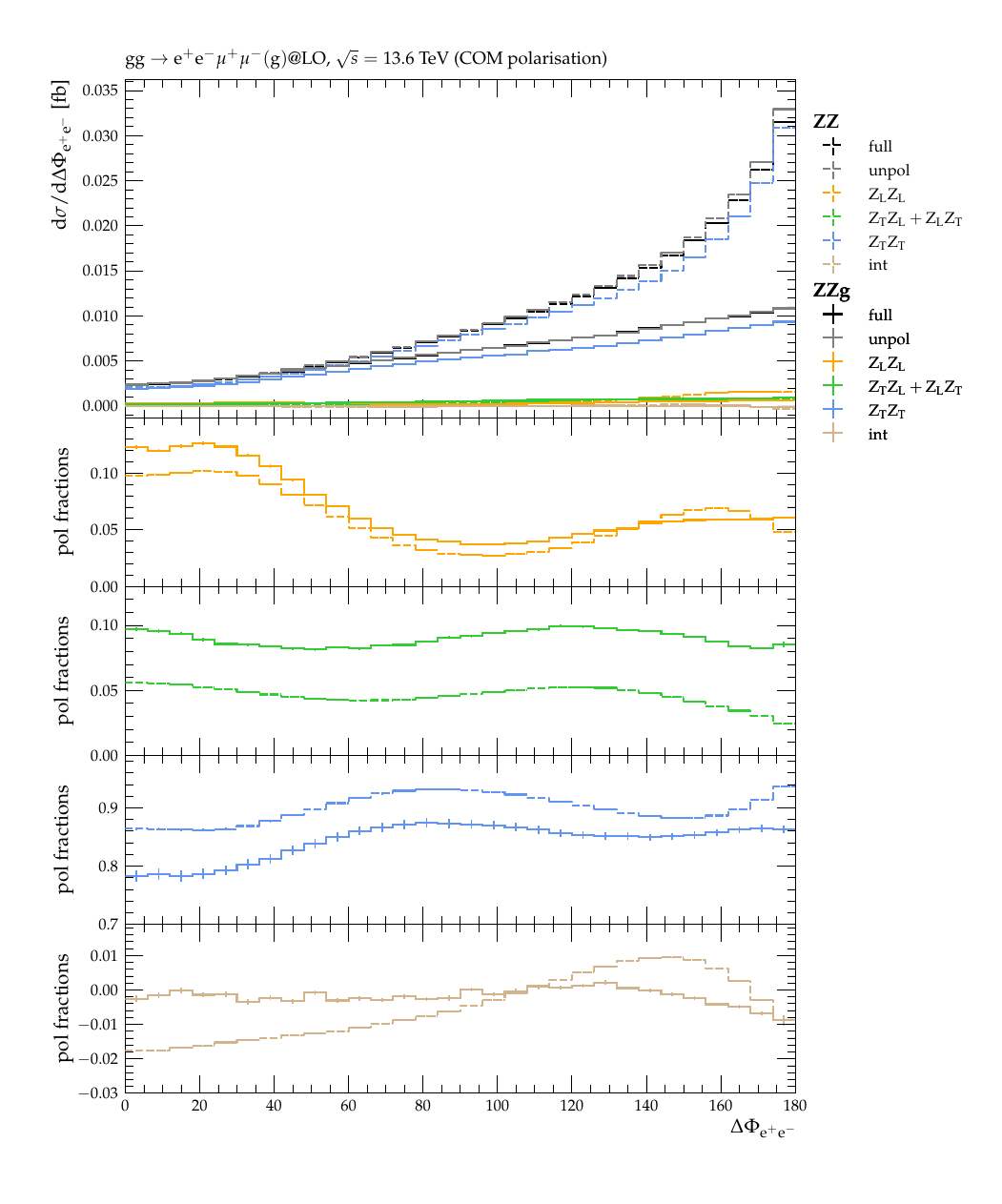}
		\end{subfigure}
	\end{subfigure}
	\caption{Polarised differential cross sections of the $\PZ\PZ$ invariant mass $\Delta \Phi_{\Pe^+\Pe^-}$ for loop-induced, inclusive and jet-associated $\PZ\PZ$ boson production at LO. Polarisation is calculated in the laboratory frame~(lab, left) and in the diboson centre-of-mass frame~(com, right). Upper panels display the absolute distributions, lower panels show the differential polarisation fractions with respect to the unpolarised result of the respective process.}
	\label{Fig.: differential XS ZZ ZZj angle}
\end{figure}

\begin{figure}[t]
	\centering
	\begin{subfigure}{0.465\textwidth}
		\begin{subfigure}{\textwidth}
			\includegraphics[width=\textwidth]{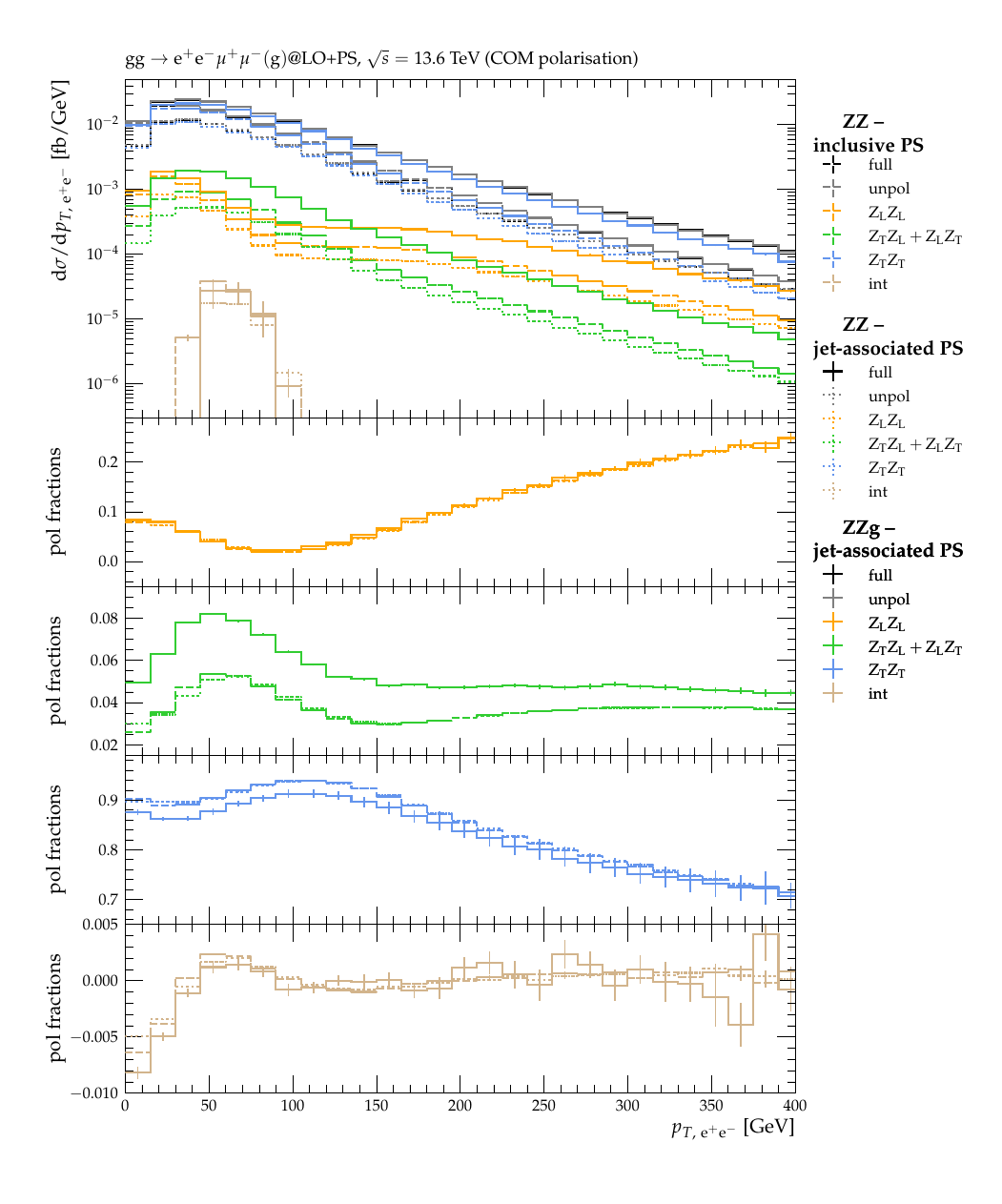}
		\end{subfigure}
	\end{subfigure}
	\hspace*{0.05\textwidth}
	\begin{subfigure}{0.465\textwidth}
		\begin{subfigure}{\textwidth}
			\includegraphics[width=\textwidth]{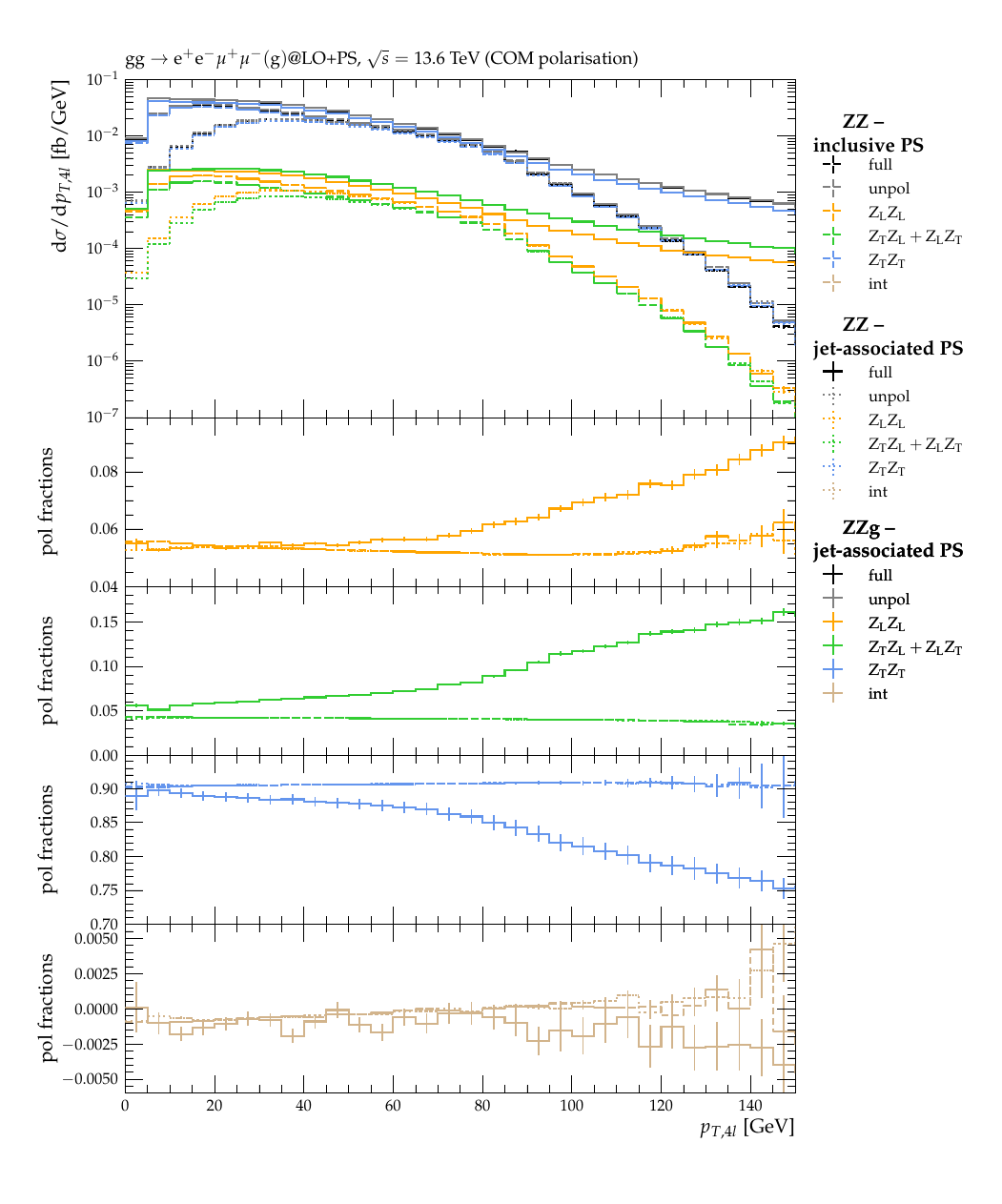}
		\end{subfigure}
	\end{subfigure}
	\caption{Polarised differential cross sections of the transverse momentum of electron and positron $p_{T,\Pe^+\Pe^-}$~(left) and of all four final-state leptons $p_{T,4l}$~(right) for loop-induced, inclusive and jet-associated $\PZ\PZ$ boson production at LO+CSS(QCD)+YFS(QED). The inclusive process is considered in both the jet-inclusive~(dashed) as well as the jet-associated phase space~(PS, dotted). Polarisation is calculated in the diboson centre-of-mass~(com) frame. Upper panels display the absolute distributions, lower panels show the differential polarisation fractions with respect to the unpolarised result of the respective process.}
	\label{Fig.: differential XS ZZ ZZ jetsPS ZZj}
\end{figure}

Fig.~\ref{Fig.: differential XS ZZ ZZj mass} and Fig.~\ref{Fig.: differential XS ZZ ZZj angle} compare the polarisation fractions of inclusive and jet-associated $\PZ\PZ$ production in their respective phase spaces at fixed order. In Fig.~\ref{Fig.: differential XS ZZ ZZ jetsPS ZZj}, inclusive $\PZ\PZ$ production is also evaluated in the jet-associated phase space which requires the inclusion of PS effects. The latter comparison illustrates that, although the additional jet requirements reduce the cross section by about $40\%$, the polarisation decomposition remains unaffected. The largest effects are found in $p_{T,\Pe^+\Pe^-}$ in the bulk region of the distribution. Including an additional jet in the matrix element, however, can lead to significant distortions in the differential polarisation decomposition, depending on the observable. For most of the studied observables such as the $m_{4l}$ and $p_{T,\,\Pe^+\Pe^-}$ distributions shown, only certain polarisation contributions feature noteworthy changes beyond an overall shift of the integrated polarisation fraction. In both cases, the mixed contribution differs most between the inclusive and jet-associated processes. For $m_{4l}$, where distributions for Lab and COM frame definitions are displayed, the difference is much more pronounced in the COM with differences of up to 10~percentage points in the polarisation fractions. This is partly compensated by a smaller TT contribution. Also, the interference contribution exhibits deviations between the two processes for small $m_{4l}$ where the inclusive process has the larger interference fraction.

Large shape distortions in all fraction distributions are observed as a function of $\Delta \Phi_{\Pe^+\Pe^-}$ and $p_{T,4l}$. In $p_{T,4l}$, they are extraordinarily large, because, for the inclusive process, non-zero $p_{T,4l}$ is generated entirely by the PS which does not change the polarisation of the intermediate boson.  

\subsection{Differential results for loop-induced inclusive $W^+W^-$ production}
\begin{figure}[!h]
	\centering
	\begin{subfigure}{0.465\textwidth}
		\begin{subfigure}{\textwidth}
			\includegraphics[width=\textwidth]{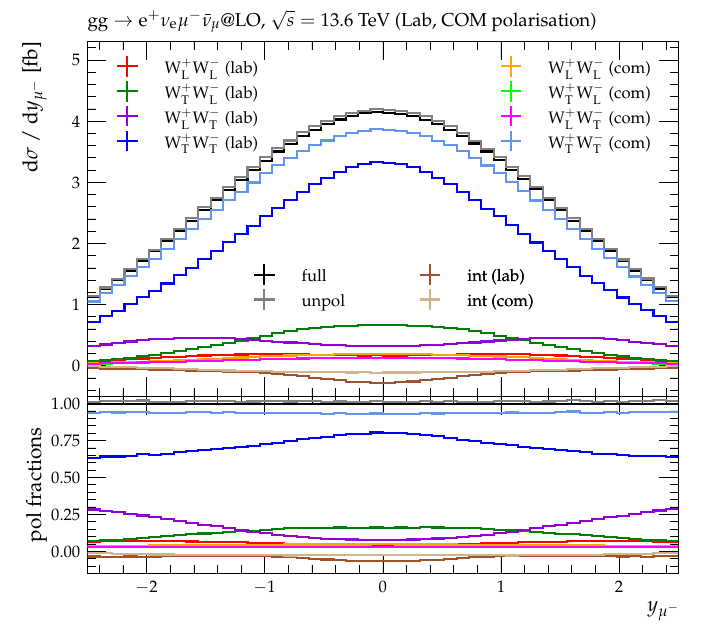}
		\end{subfigure}
	\end{subfigure}
	\hspace*{0.025\textwidth}
	\begin{subfigure}{0.465\textwidth}
		\begin{subfigure}{\textwidth}
			\includegraphics[width=\textwidth]{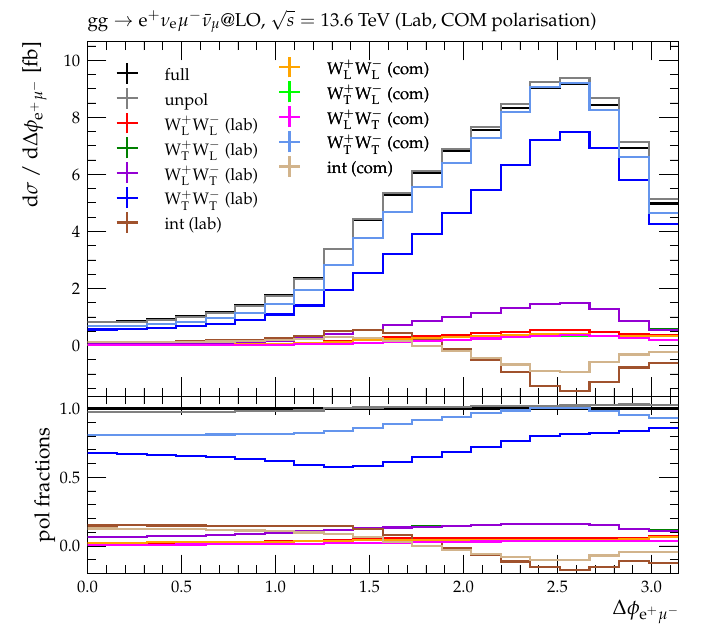}
		\end{subfigure}
	\end{subfigure}
	\caption{Polarised differential cross sections of the muon rapidity $y_{\mu^-}$~(left) and the azimuthal angle difference between positron and muon $\Delta \Phi_{\mathrm{e}^+ \mu^-}$~(right) for loop-induced, inclusive $\PW^+\PW^-$ boson production at LO. Polarisation is calculated in the laboratory~(lab) and the diboson centre-of-mass~(com) frame. Upper panels display the absolute distributions, lower panels show the differential polarisation fractions relative to the unpolarised result.}
	\label{Fig.: differential XS WW FO}
\end{figure}

\begin{figure}[!h]
	\centering
	\begin{subfigure}{0.465\textwidth}
		\begin{subfigure}{\textwidth}
			\includegraphics[width=\textwidth]{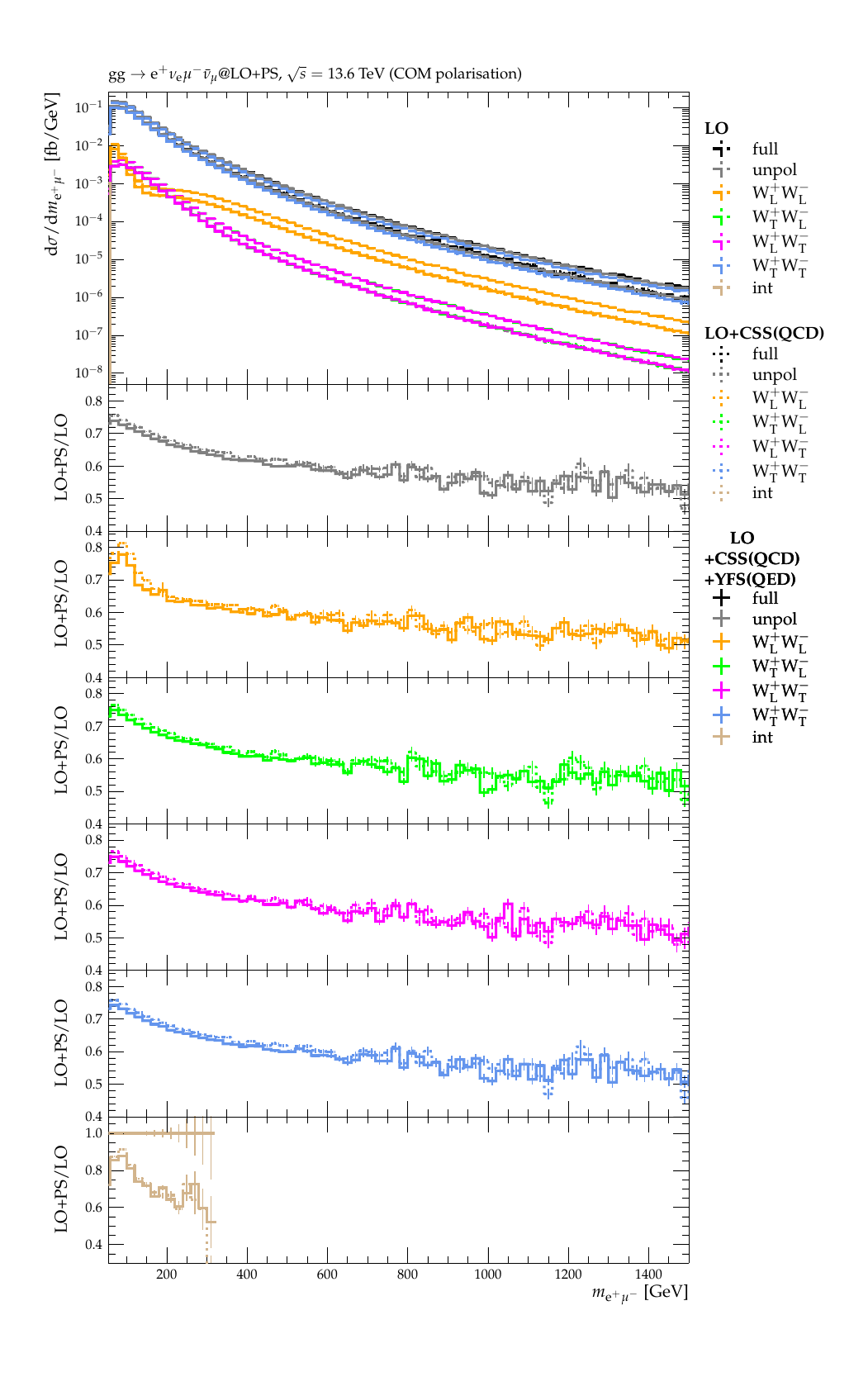}
		\end{subfigure}
	\end{subfigure}
	\hspace*{0.025\textwidth}
	\begin{subfigure}{0.465\textwidth}
		\begin{subfigure}{\textwidth}
			\includegraphics[width=\textwidth]{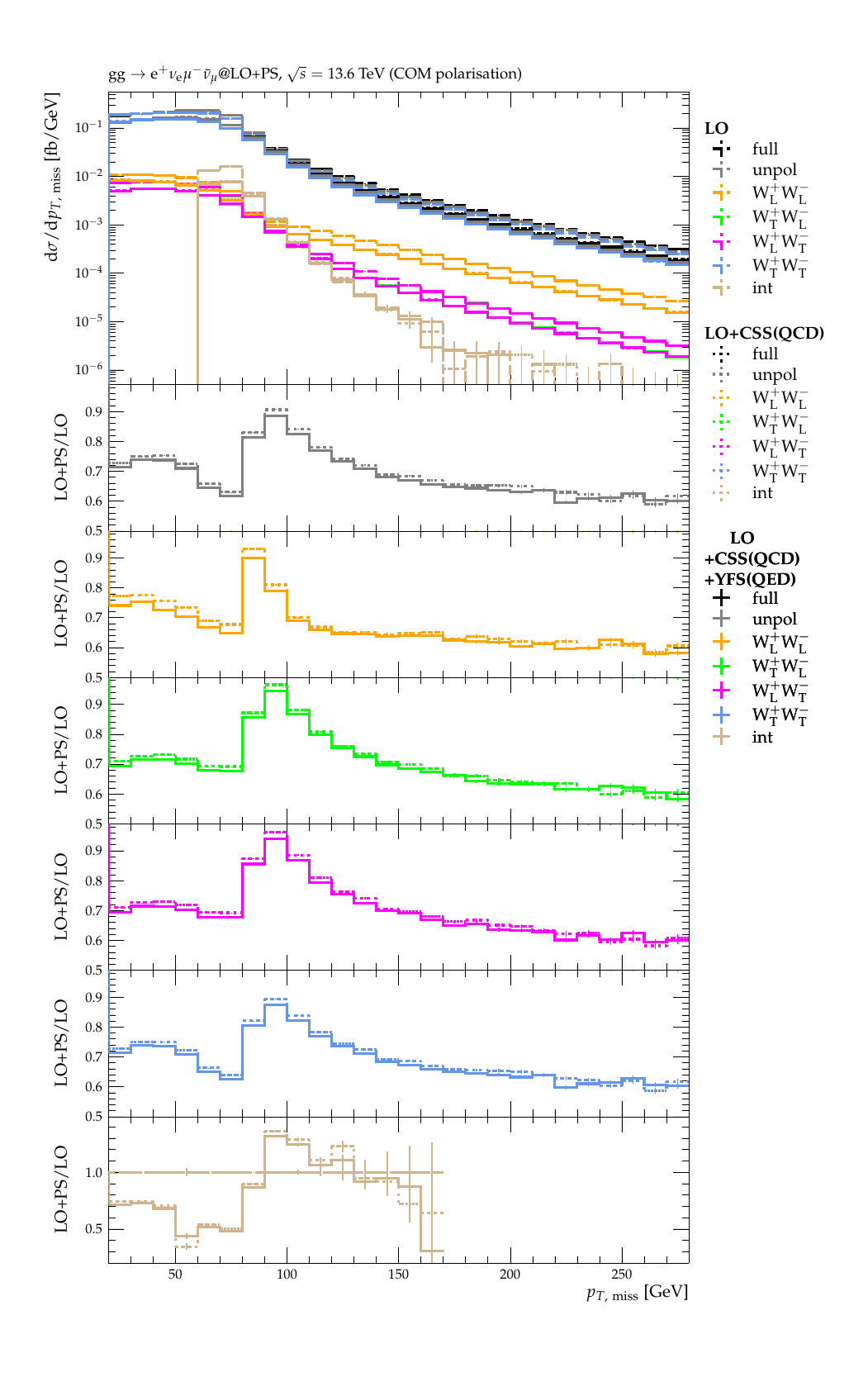}
		\end{subfigure}
	\end{subfigure}
	\caption{Polarised differential cross sections of the invariant mass of the positron--muon system $m_{\Pe^+\mu^-}$~(left) and the missing transverse momentum $p_{T,miss}$~(right) for loop-induced, inclusive $\PW^+\PW^-$ boson production at LO, LO matched to a QCD shower only (LO+CSS(QCD)) and LO matched to QCD and QED resummation (LO+CSS(QCD)+YFS(QED)). Polarisation is calculated in the diboson centre-of-mass~(com) frame. Upper panels display the absolute distributions, lower panels show the cross sections including PS effects relative to their fixed-order counterparts for the on-shell unpolarised calculation and the different polarisation contributions. If not shown, interference contributions are zero in that region.}
	\label{Fig.: differential XS WW LOPS}
\end{figure}

For loop-induced $\PW^+\PW^-$ production, differential distributions in the muon rapidity $y_{\mu^-}$ and of the azimuthal separation between the charged leptons $\Delta \Phi_{\mathrm{e}^+ \mu^-}$ at fixed LO are shown in Fig.~\ref{Fig.: differential XS WW FO}. 
Fig.~\ref{Fig.: differential XS WW LOPS} exhibits PS effects in the invariant mass of the two charged leptons $m_{\Pe^+\mu^-}$ and the missing-momentum $p_{T,\text{miss}}$ distributions.  

Off-shell effects are at the percent level in all displayed distributions. In the polarised distributions, the transverse polarisation contribution is dominant in the entire phase space, similar to $\PZ\PZ$ production. The mixed polarisation contributions exhibit a markedly different behaviour depending on the observable. In the $y_{\mu^-}$ distributions in the Lab frame definition, the TL component resembles the TT and unpolarised distributions, with a single maximum around zero, whereas the LT component is distinctly different, showing a minimum at $y_{\mu^-}=0.0$ and two symmetric maxima at approximately $y_{\mu^-}=\pm 1.5$. In contrast, no appreciable differences among the mixed contributions are observed in the COM definition, nor in $\Delta \Phi_{\mathrm{e}^+ \mu^-}$ or in any of the energy- and momentum-related observables studied. 

The longitudinal contribution is subdominant, but exceeds the mixed contributions over broad regions in the COM frame definition. In the $y_{\mu^-}$ distribution evaluated in the Lab frame, the LL cross section exhibits a shape similar to the LT distribution, while in the COM frame it peaks at zero, in line with all other polarisation contributions. In the $\Delta \Phi_{\Pe^+ \mu^-}$ distribution, the LL contribution is largest in the bulk of the cross section where all components peak. 

The interference contributions become significant in certain regions of phase space, reaching magnitudes comparable to, or in some regions larger than, those of the subleading pure-polarisation components. In particular, in $\Delta \Phi_{\mathrm{e}^+ \mu^-}$, the interference contribution reaches nearly 10\% over broad regions of the phase space, becoming the second-dominant contribution. 

The choice of reference frame affects not only the longitudinally polarised components but also the interference and TT contributions. This is especially apparent in the $y_{\mu^-}$ distribution, where both the TT and interference fraction contributions are nearly flat in the COM definition but acquire a pronounced shape dependence in the Lab frame. 

Resummation effects on the polarisation fractions remain rather small, owing to the strict jet selection criteria that preferentially select fixed-order-like kinematic configurations. Among the pure-polarisation components, visible shape changes occur only for the longitudinal contribution. These modifications are compensated almost entirely by corresponding changes in the interference distribution.
 
\subsection{Differential results for loop-induced inclusive $Z\gamma$ production}
\begin{figure}[htbp]
	\centering
	\begin{subfigure}{0.465\textwidth}
		\begin{subfigure}{\textwidth}
			\includegraphics[width=\textwidth]{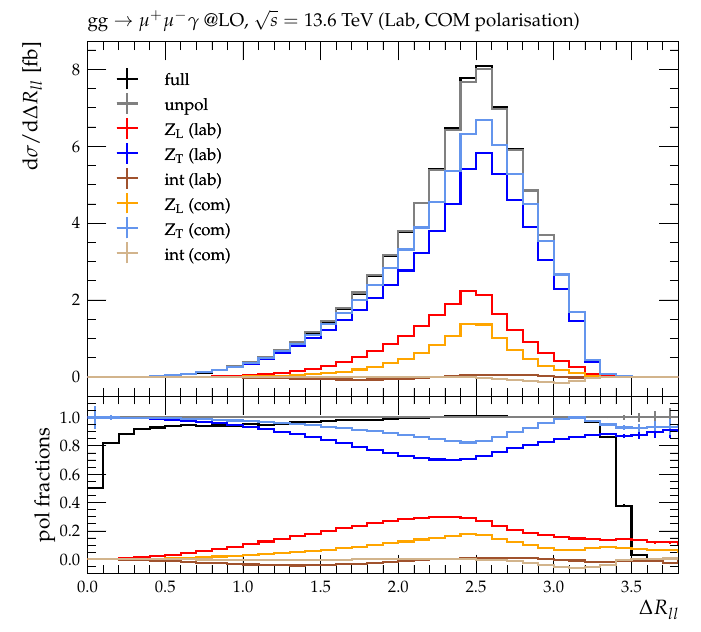}
		\end{subfigure}
	\end{subfigure}
	\hspace*{0.05\textwidth}
	\begin{subfigure}{0.465\textwidth}
		\begin{subfigure}{\textwidth}
			\includegraphics[width=\textwidth]{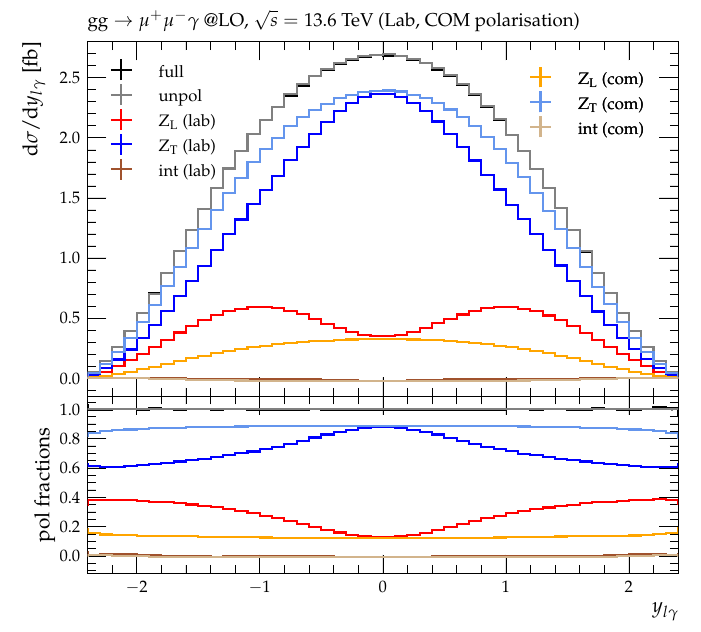}
		\end{subfigure}
	\end{subfigure}
	\caption{Polarised differential cross sections of the dilepton angular separation $\Delta R_{ll}$~(left) and the rapidity of the positron-photon-system $y_{l\gamma}$~(right) for loop-induced, inclusive $\PZ\gamma$ boson production at LO. Polarisation is calculated in the laboratory~(lab) and in the diboson centre-of-mass~(com) frame. Upper panels display the absolute distributions, lower panels show the differential polarisation fractions relative to the unpolarised result.}
	\label{Fig.: differential XS Zy FO}
\end{figure}

\begin{figure}[htbp]
	\centering
	\begin{subfigure}{0.465\textwidth}
		\begin{subfigure}{\textwidth}
			\includegraphics[width=\textwidth]{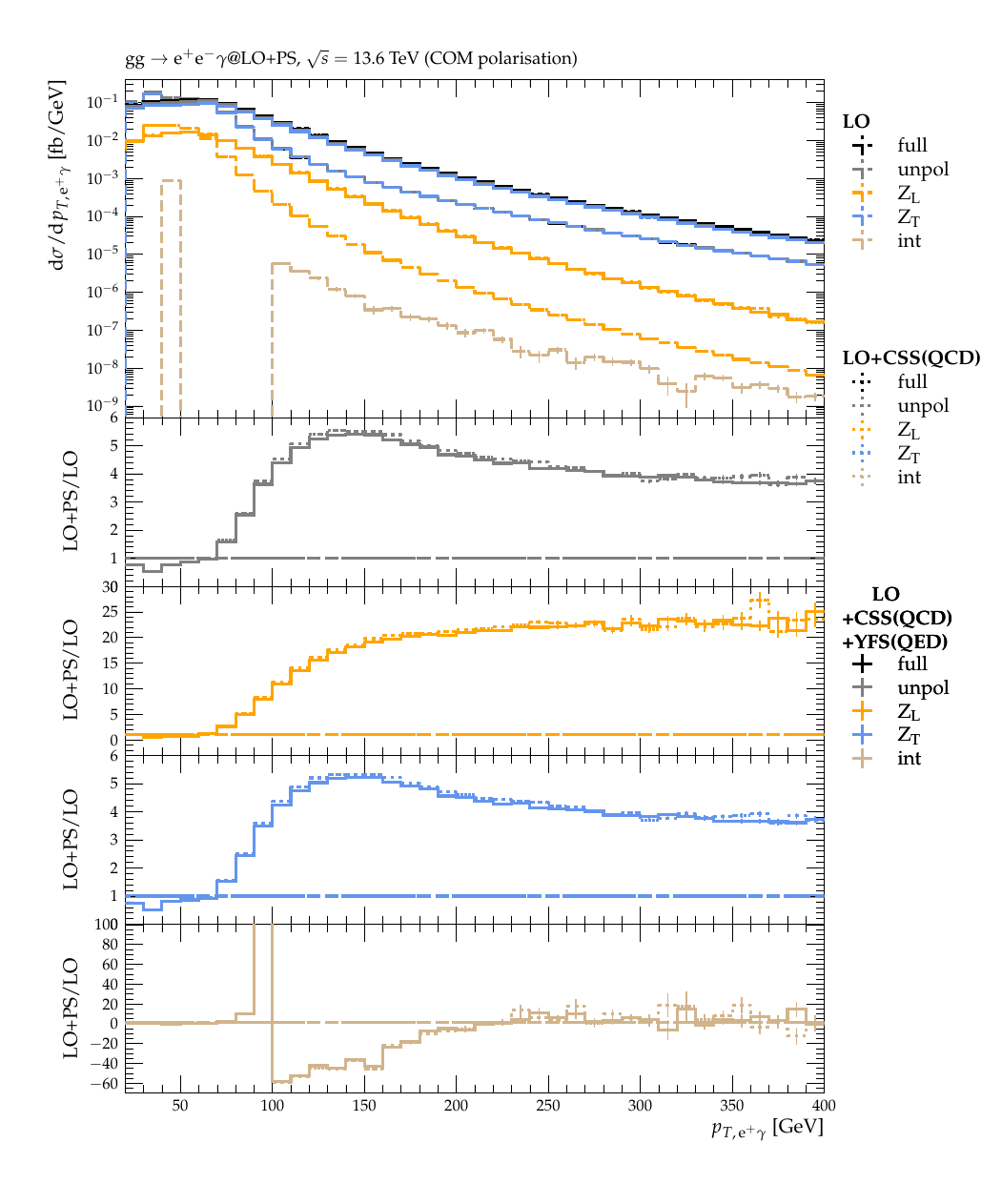}
		\end{subfigure}
	\end{subfigure}
	\hspace*{0.05\textwidth}
	\begin{subfigure}{0.465\textwidth}
		\begin{subfigure}{\textwidth}
			\includegraphics[width=\textwidth]{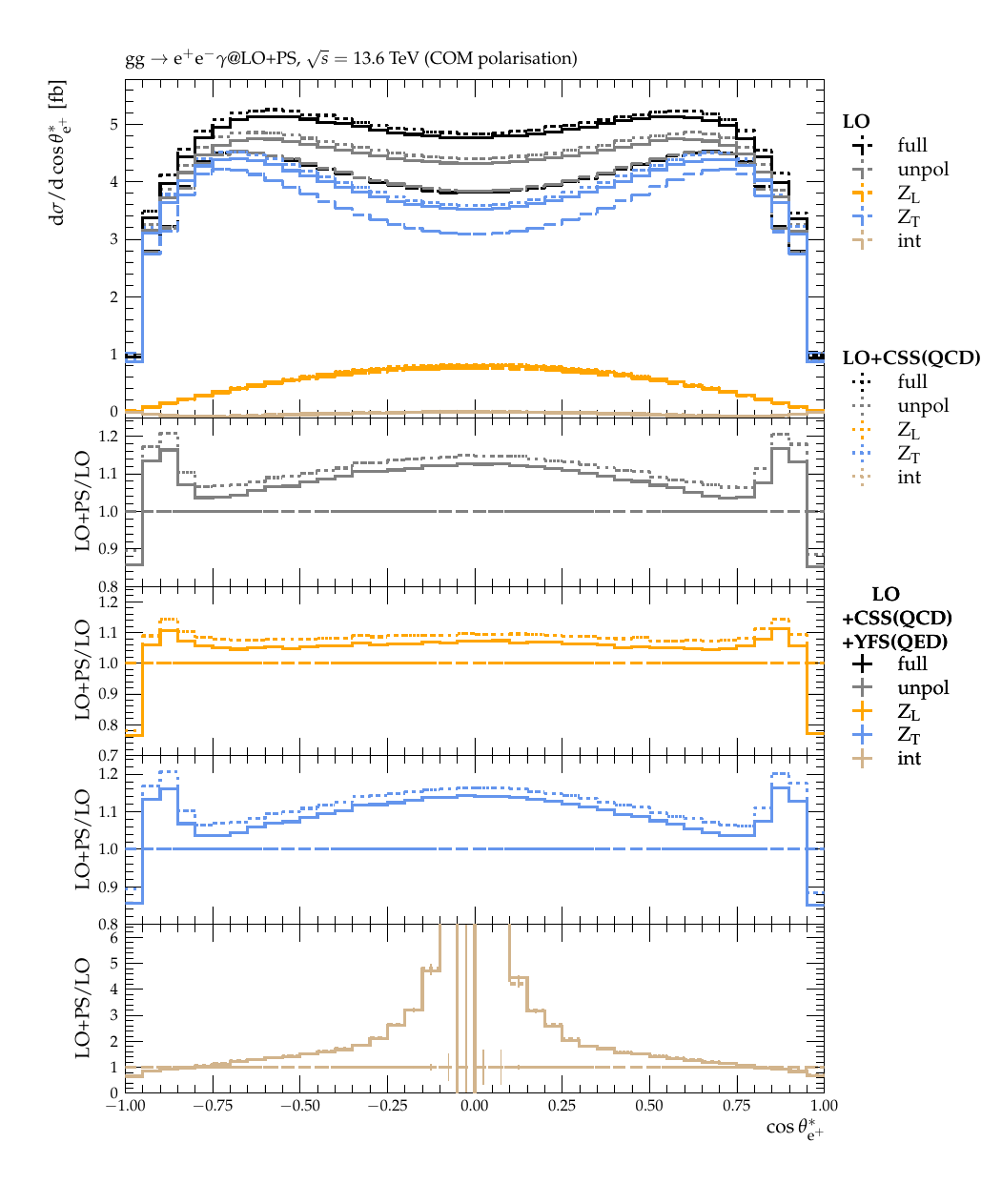}
		\end{subfigure}
	\end{subfigure}
	\caption{Polarised differential cross sections of the transverse momentum of positron and photon $p_{T,\Pe^+ \gamma}$~(left) and of the lepton decay angle of the positron $\cos \theta^*_{\Pe^+}$~(right) for loop-induced, inclusive $\PZ\gamma$ boson production at LO, LO matched to a QCD shower only (LO+CSS(QCD)) and LO matched to QCD and QED resummation (LO+CSS(QCD)+YFS(QED)). Polarisation is calculated in the diboson centre-of-mass~(com) frame.  Upper panels display the absolute distributions, lower panels show the cross sections including PS effects relative to their fixed-order counterparts for the on-shell unpolarised calculation and the different polarisation contributions.} 
	\label{Fig.: differential XS Zy LOPS}
\end{figure}

Fig.~\ref{Fig.: differential XS Zy FO} displays the dilepton angular separation $\Delta R_{ll}$ and the rapidity of the positron-photon-system $y_{l\gamma}$ at fixed order, calculated for loop-induced, inclusive $\PZ\gamma$ production. The impact of resummation effects on the transverse momentum of positron and photon $p_{T, \Pe^+ \gamma}$ and the positron decay angle $\cos \theta^*_{\Pe^+}$ distributions is shown in Fig.~\ref{Fig.: differential XS Zy LOPS}. 

In this process, the level of agreement between the on-shell and off-shell calculations depends strongly on the observable. In $y_{l\gamma}$, the discrepancy is negligible, whereas $\Delta R_{ll}$ shows large differences in suppressed phase-space regions: the off-shell distribution falls off substantially faster at both large and small separations compared to the on-shell result. When PS effects are included, the differences between on- and off-shell results increase substantially.

Similar to the previously discussed processes, the transverse contribution remains dominant across the entire phase space. For $\Delta R_{ll}$, beyond the overall shift of the integrated fractions, different polarisation definitions produce slightly different shapes around the maximum of the distribution and slightly displaced maxima of the fraction distributions. In $y_{l\gamma}$, on the other hand, substantial shape differences emerge. The transverse cross-section distribution is considerably broader and closer to the unpolarised distribution in the COM frame, resulting in a nearly flat fraction distribution. In contrast, in the Lab frame, the transverse fraction peaks at $y_{l\gamma}=0$ and then decreases from approximately $82\%$ to about $60\%$ at large rapidities. The longitudinal cross section shows two maxima around $y_{l\gamma}=\pm 1$ in the Lab frame, while in the COM frame it peaks at zero and, much like the transverse distribution, closely matches the unpolarised shape, yielding a flat polarisation fraction. Consequently, this observable is sensitive to the different polarisation states only when evaluated in the Lab frame. 

A pronounced interference structure appears only in $\Delta R_{ll}$, where the fractions reach at most $\pm5\%$. 

Resummation effects are again dominated by QCD corrections and affect exclusively the longitudinal and interference fractions in the distributions shown. In the tail of the $p_{T \Pe^+ \gamma}$ distribution, the longitudinal cross section is enhanced by up to a factor of 25 compared to fixed order, while the T and unpolarised cross sections increase by at most a factor of five. The enhanced longitudinal fraction is compensated by the interference component, which, in contrast to the strictly positive interference across the entire tail at fixed order, becomes strongly negative in the range $100 \gev \leq p_{T \Pe^+ \gamma} \leq 200\gev$. In the $\cos \theta^*_{\Pe^+}$ distribution, the longitudinal resummed cross section deviates from the fixed-order result more uniformly over the phase space than the T and unpolarised contributions. This non-trivial change in the L contribution due to resummation effects is primarily compensated by an increase in the interference fraction centred around $\cos \theta^*_{\Pe^+}=0$. Additional QED radiation mainly produces an overall reduction of the cross section, with no significant impact on the distribution shapes or polarisation fractions.

\subsection{Differential results for loop-induced, inclusive $Zh$ production}
\begin{figure}[t]
	\centering
	\begin{subfigure}{0.465\textwidth}
		\begin{subfigure}{\textwidth}
			\includegraphics[width=\textwidth]{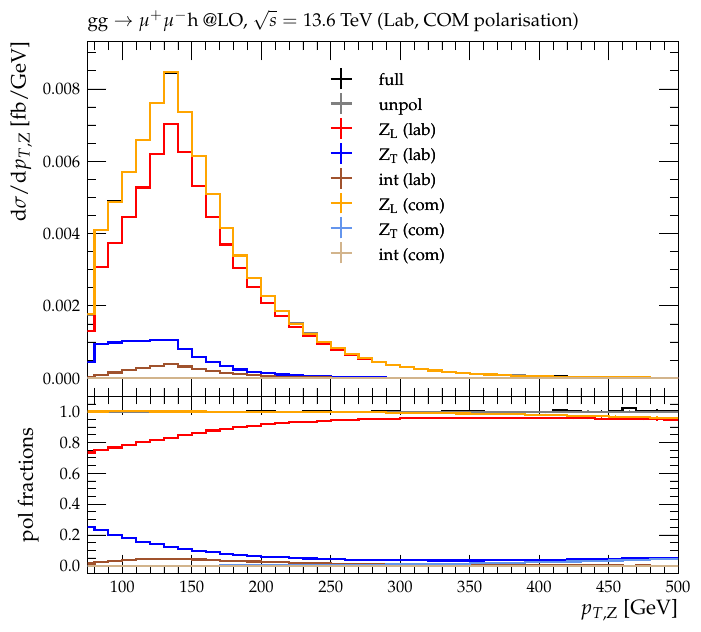}
		\end{subfigure}
	\end{subfigure}
	\hspace*{0.05\textwidth}
	\begin{subfigure}{0.465\textwidth}
		\begin{subfigure}{\textwidth}
			\includegraphics[width=\textwidth]{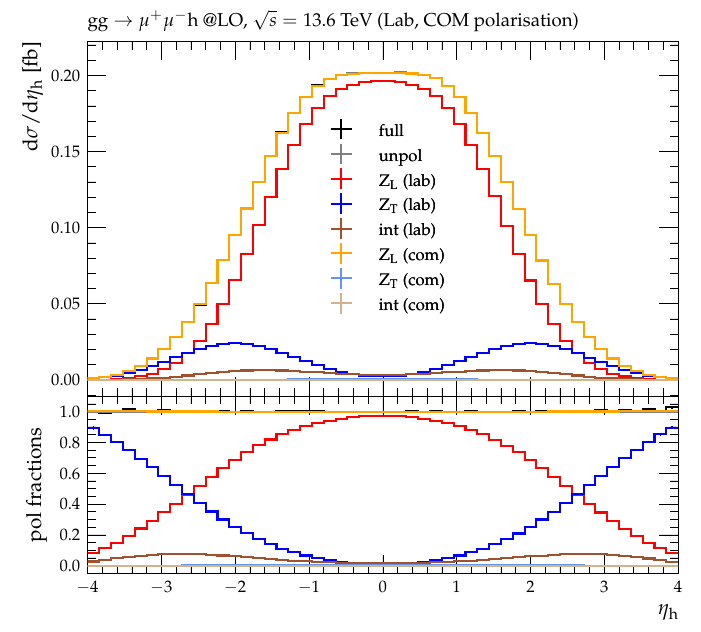}
		\end{subfigure}
	\end{subfigure}
	\caption{Polarised differential cross sections of the transverse momentum of the $\PZ$ boson $p_{T, \PZ}$~(left) and the pseudorapidity of the Higgs boson $\eta_{h}$~(right) for loop-induced, inclusive $\PZ\PH$ boson production at LO. Polarisation is calculated in the laboratory~(lab) and in the diboson centre-of-mass~(com) frame. Upper panels display the absolute distributions, lower panels show the differential polarisation fractions relative to the unpolarised result.  }
	\label{Fig.: differential XS Zh FO}
\end{figure}

\begin{figure}[!t]
	\centering
	\begin{subfigure}{0.49\textwidth}
		\begin{subfigure}{\textwidth}
			\includegraphics[width=\textwidth, trim=0cm 1cm 0cm 0.8cm, clip]{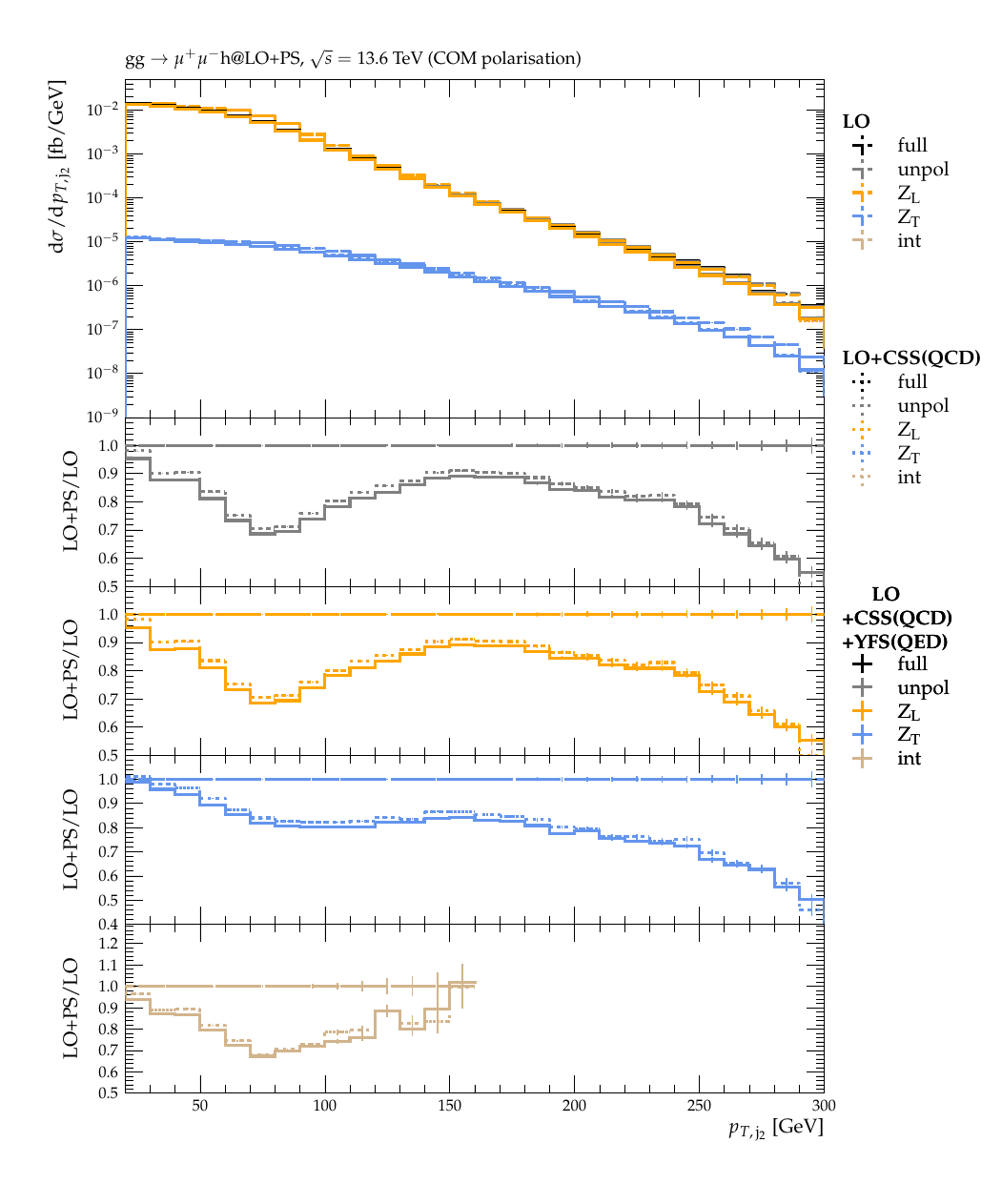}
		\end{subfigure}
	\end{subfigure}
	\begin{subfigure}{0.49\textwidth}
		\begin{subfigure}{\textwidth}
			\includegraphics[width=\textwidth, trim=0cm 1cm 0cm 0.8cm, clip]{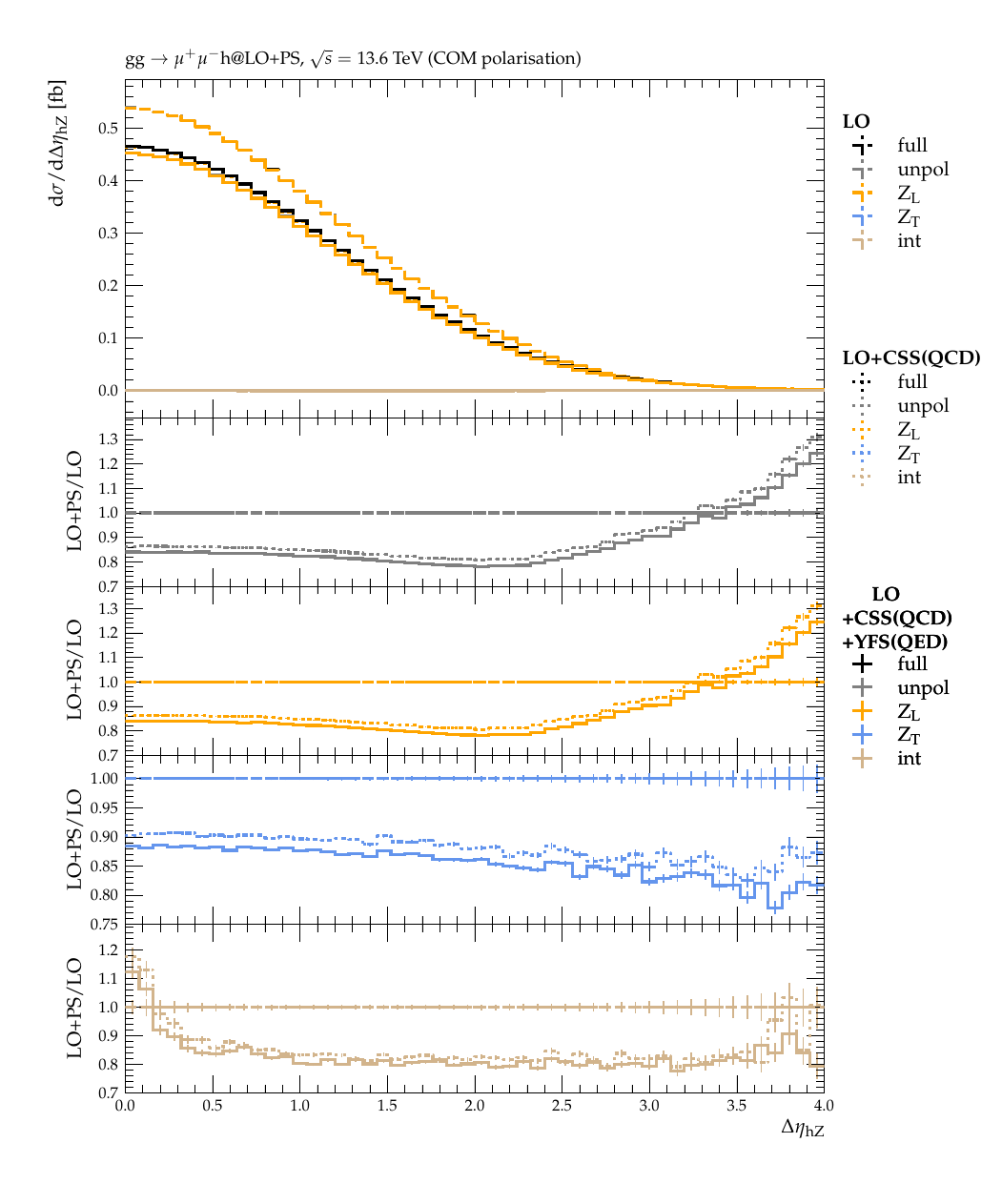}
		\end{subfigure}
	\end{subfigure}
	\caption{Polarised differential cross sections of the transverse momentum of the subleading b-jet from the Higgs boson decay $p_{T,\Pj_2}$~(left) and the pseudorapidity difference between Higgs and the $\PZ$ boson $\Delta\eta_{\PH\PZ}$~(right) for loop-induced, inclusive $\PZ\PH$ boson production at fixed \lo, LO matched to a QCD shower only (LO+CSS(QCD)) and LO matched to QCD and QED resummation (LO+CSS(QCD)+YFS(QED)). Polarisation is defined in the diboson centre-of-mass~(com) frame.  Upper panels display the absolute distributions, lower panels show the cross sections including parton-shower effects relative to their fixed-order counterparts for the on-shell unpolarised calculation and the different polarisation contributions.} 
	\label{Fig.: differential XS Zh LOPS}
\end{figure}

Differential cross sections in the transverse momentum of the $\PZ$ boson $p_{T, \PZ}$ and the pseudorapidity of the Higgs boson $\eta_\PH$ for inclusive, loop-induced $\PZ\PH$ production at fixed LO are presented in Fig.~\ref{Fig.: differential XS Zh FO}. PS-matched differential distributions of the transverse momentum of the subleading jet from the $\PH \rightarrow \Pb \bar{\Pb}$ decay $p_{T,\, \Pj_2}$ and of the pseudorapidity separation between the Higgs and $\PZ$ bosons $\Delta \eta_{hZ}$ are shown in Fig.~\ref{Fig.: differential XS Zh LOPS}. 

Both at fixed order and after including only QCD PS effects, off-shell effects remain negligible, also at the differential level. When QED radiation is added, larger differences between the off-shell and on-shell calculations are observed. The longitudinal polarisation is the dominant contribution in this process. Only in the Lab-frame definition do non-negligible transverse contributions appear over broad regions of phase space.  For the COM frame definition, a significant T contribution only appears for high values of $p_{T,\PZ}$. In this region of phase space, both polarisation definitions coincide.  

The transverse polarisation exhibits a distinctly different shape compared to the longitudinal contribution in both observables shown. Unlike the longitudinal/unpolarised component in the Lab frame, the transverse $p_{T,\PZ}$ cross section distribution has a much broader maximum, and the transverse $\eta_\PH$ distribution features two maxima around $\eta_\PH=\pm 2$ instead of a single peak centred at $\eta_\PH = 0$. The transverse polarisation fraction in the Lab frame is therefore largest at small $p_{T,\PZ}$, reaching up to $25\%$, and at large $\PH/\PZ$ boson rapidities, where it becomes the dominant polarisation contribution.


The interference contribution is likewise sizeable only in the Lab frame definition, reaching at most $5\%$. 
Its cross section (fraction) distribution displays a shape distinct from the other contributions, with an almost symmetric $p_{T,\PZ}$ distribution centred around $p_{T,\PZ}=135$~GeV, as well as very broad maxima at $\eta_\PH =1.5~(2.7)$. 

QCD radiation again drives trivial distortion of the fixed-order distributions through resummation effects. The deviation from the fixed-order result is flatter for the transverse than for the longitudinal contribution, although the transverse contribution remains numerically small. The interference largely follows the unpolarised distribution in $p_{T,\, \Pj_2}$, while in $\Delta \eta_{\PH\PZ}$ it shows a uniform reduction over most of the phase space compared to fixed order. This is in contrast to the longitudinal and the unpolarised cross section, which exceed the fixed-order result at large rapidity separations. At small rapidity separations, the interference exhibits a larger cross section than the fixed-order prediction, unlike all other contributions.
\section{Conclusions}
\label{sec:conclusions}

In this work, the simulation of polarised processes in the \Sherpa Monte Carlo event generator has been extended to loop-induced processes. Simulations at fixed order and matched to a parton shower are presented for a range of loop-induced processes, several studied for the first time in this form. Motivated by the increasing importance of polarisation measurements as probes of the electroweak symmetry-breaking mechanism and the gauge sector of the Standard Model, this development addresses a critical gap in the precision modelling required for current and future LHC analyses, particularly in the high-luminosity phase.

The new implementation, based on polarised amplitudes from the RECOLA matrix element generator, has been validated against existing results, demonstrating excellent agreement for benchmark processes such as $gg \to W^+W^-$. Detailed studies of polarised cross sections have been presented for several loop-induced processes, including $gg \to W^+W^-$, $gg\to ZZ$, $gg\to Z\gamma$, and $gg\to Zh$, as well as their jet-associated counterparts. Novel results for many of the jet-associated processes allow for the first time to judge higher-order effects on the polarisation decomposition in these loop-induced processes. 

The presented calculations have been performed in two different frames that significantly influence the polarisation decomposition, as known from previous studies of
different processes. Resummation effects can distort the polarisation decomposition significantly compared to the fixed-order result and thus make their inclusion in realistic polarisation templates imperative. This is especially evident for jet-associated processes where resummation effects are even visible at the level of fiducial polarisation fractions. For inclusive processes, changes in the fiducial fractions remain below 1\% but differential distributions, especially of the interesting longitudinal components, are also distorted substantially.

Looking ahead, several avenues for further development are apparent. The availability of jet-associated loop-induced processes with parton showering will enable in a next step the generation of multijet-merged samples to provide higher-order predictions for the inclusive processes, achieving the precision required by upcoming LHC measurements. The public release of the polarisation framework and its integration with experimental workflows will facilitate its application to real LHC data, enhancing the sensitivity of polarisation measurements in the HL-LHC era.

\section*{Acknowledgements}

MS is funded by the Royal Society through a University Research Fellowship
(URF\textbackslash{}R1\textbackslash{}180549, URF\textbackslash{}R\textbackslash{}231031) and an Enhancement Award
(RGF\textbackslash{}EA\textbackslash{}181033,
 CEC19\textbackslash{}100349, and RF\textbackslash{}ERE\textbackslash{}210397)
as well as the STFC (ST/X003167/1 and ST/X000745/1). 
MH was funded by a COMETA COST Action STSM Grant No. 29f14d41 and thanks the Durham University for hosting her to work on this project. Furthermore, MH acknowledges support by a doctoral scholarship from the Friedrich-Ebert-Stiftung during finalising steps of this work. 

The authors acknowledge support from the COMETA COST Action CA22130 and thank Giovanni Pelliccioli for useful discussions regarding Recola. 

\bibliographystyle{amsunsrt_mod}
\bibliography{refs}

@article{Sherpa:2024mfk,
    author = "Bothmann, Enrico and others",
    collaboration = "Sherpa",
    title = "{Event generation with Sherpa 3}",
    eprint = "2410.22148",
    archivePrefix = "arXiv",
    primaryClass = "hep-ph",
    reportNumber = "IPPP/24/67, LTH-1385, FERMILAB-PUB-24-0748-T, ZU-TH 51/24, MCNET-24-17, CERN-TH-2024-171",
    doi = "10.1007/JHEP12(2024)156",
    journal = "JHEP",
    volume = "12",
    pages = "156",
    year = "2024"
}

@article{Sherpa:2019gpd,
    author = "Bothmann, Enrico and others",
    collaboration = "Sherpa",
    title = "{Event Generation with Sherpa 2.2}",
    eprint = "1905.09127",
    archivePrefix = "arXiv",
    primaryClass = "hep-ph",
    reportNumber = "FERMILAB-PUB-19-218-T, SLAC-PUB-17433, IPPP/19/42, MCNET-19-11",
    doi = "10.21468/SciPostPhys.7.3.034",
    journal = "SciPost Phys.",
    volume = "7",
    number = "3",
    pages = "034",
    year = "2019"
}

@article{Bierlich:2022pfr,
    author = "Bierlich, Christian and others",
    title = "{A comprehensive guide to the physics and usage of PYTHIA 8.3}",
    eprint = "2203.11601",
    archivePrefix = "arXiv",
    primaryClass = "hep-ph",
    reportNumber = "LU-TP 22-16, MCNET-22-04, FERMILAB-PUB-22-227-SCD",
    doi = "10.21468/SciPostPhysCodeb.8",
    journal = "SciPost Phys. Codeb.",
    volume = "2022",
    pages = "8",
    year = "2022"
}

@article{Bellm:2015jjp,
    author = "Bellm, Johannes and others",
    title = "{Herwig 7.0/Herwig++ 3.0 release note}",
    eprint = "1512.01178",
    archivePrefix = "arXiv",
    primaryClass = "hep-ph",
    reportNumber = "CERN-PH-TH-2015-289, MAN-HEP-2015-15, IFJPAN-IV-2015-13, KA-TP-18-2015, DCPT-15-142, MCNET-15-28, IPPP-15-71, HERWIG-2015-01",
    doi = "10.1140/epjc/s10052-016-4018-8",
    journal = "Eur. Phys. J. C",
    volume = "76",
    number = "4",
    pages = "196",
    year = "2016"
}

@article{Bahr:2008pv,
    author = "Bahr, M. and others",
    title = "{Herwig++ Physics and Manual}",
    eprint = "0803.0883",
    archivePrefix = "arXiv",
    primaryClass = "hep-ph",
    reportNumber = "CERN-PH-TH-2008-038, CAVENDISH-HEP-08-03, KA-TP-05-2008, DCPT-08-22, IPPP-08-11, CP3-08-05",
    doi = "10.1140/epjc/s10052-008-0798-9",
    journal = "Eur. Phys. J. C",
    volume = "58",
    pages = "639--707",
    year = "2008"
}

@article{Nason:2004rx,
    author = "Nason, Paolo",
    title = "{A New method for combining NLO QCD with shower Monte Carlo algorithms}",
    eprint = "hep-ph/0409146",
    archivePrefix = "arXiv",
    reportNumber = "BICOCCA-FT-04-11",
    doi = "10.1088/1126-6708/2004/11/040",
    journal = "JHEP",
    volume = "11",
    pages = "040",
    year = "2004"
}

@article{Frixione:2007vw,
    author = "Frixione, Stefano and Nason, Paolo and Oleari, Carlo",
    title = "{Matching NLO QCD computations with Parton Shower simulations: the POWHEG method}",
    eprint = "0709.2092",
    archivePrefix = "arXiv",
    primaryClass = "hep-ph",
    reportNumber = "BICOCCA-FT-07-9, GEF-TH-21-2007",
    doi = "10.1088/1126-6708/2007/11/070",
    journal = "JHEP",
    volume = "11",
    pages = "070",
    year = "2007"
}

@article{Alioli:2010xd,
    author = "Alioli, Simone and Nason, Paolo and Oleari, Carlo and Re, Emanuele",
    title = "{A general framework for implementing NLO calculations in shower Monte Carlo programs: the POWHEG BOX}",
    eprint = "1002.2581",
    archivePrefix = "arXiv",
    primaryClass = "hep-ph",
    reportNumber = "DESY-10-018, SFB-CPP-10-22, IPPP-10-11, DCPT-10-22",
    doi = "10.1007/JHEP06(2010)043",
    journal = "JHEP",
    volume = "06",
    pages = "043",
    year = "2010"
}

@article{Alwall:2014hca,
    author = "Alwall, J. and Frederix, R. and Frixione, S. and Hirschi, V. and Maltoni, F. and Mattelaer, O. and Shao, H. -S. and Stelzer, T. and Torrielli, P. and Zaro, M.",
    title = "{The automated computation of tree-level and next-to-leading order differential cross sections, and their matching to parton shower simulations}",
    eprint = "1405.0301",
    archivePrefix = "arXiv",
    primaryClass = "hep-ph",
    reportNumber = "CERN-PH-TH-2014-064, CP3-14-18, LPN14-066, MCNET-14-09, ZU-TH-14-14",
    doi = "10.1007/JHEP07(2014)079",
    journal = "JHEP",
    volume = "07",
    pages = "079",
    year = "2014"
}

@article{Ballestrero:2007xq,
    author = "Ballestrero, Alessandro and Belhouari, Aissa and Bevilacqua, Giuseppe and Kashkan, Vladimir and Maina, Ezio",
    title = "{PHANTOM: A Monte Carlo event generator for six parton final states at high energy colliders}",
    eprint = "0801.3359",
    archivePrefix = "arXiv",
    primaryClass = "hep-ph",
    reportNumber = "DFTT-13-2007",
    doi = "10.1016/j.cpc.2008.10.005",
    journal = "Comput. Phys. Commun.",
    volume = "180",
    pages = "401--417",
    year = "2009"
}

@article{Ballestrero:2017bxn,
    author = "Ballestrero, Alessandro and Maina, Ezio and Pelliccioli, Giovanni",
    title = "{$W$ boson polarization in vector boson scattering at the LHC}",
    eprint = "1710.09339",
    archivePrefix = "arXiv",
    primaryClass = "hep-ph",
    reportNumber = "VBSCAN-PUB-02-17",
    doi = "10.1007/JHEP03(2018)170",
    journal = "JHEP",
    volume = "03",
    pages = "170",
    year = "2018"
}

@article{BuarqueFranzosi:2019boy,
    author = "Buarque Franzosi, Diogo and Mattelaer, Olivier and Ruiz, Richard and Shil, Sujay",
    title = "{Automated predictions from polarized matrix elements}",
    eprint = "1912.01725",
    archivePrefix = "arXiv",
    primaryClass = "hep-ph",
    reportNumber = "CP3-19-52, VBSCan-PUB-09-19, MCnet-19-24, IP/BBSR/2019-10",
    doi = "10.1007/JHEP04(2020)082",
    journal = "JHEP",
    volume = "04",
    pages = "082",
    year = "2020"
}

@article{Javurkova:2024bwa,
    author = "Javurkova, Martina and Ruiz, Richard and de S\'a, Rafael Coelho Lopes and Sandesara, Jay",
    title = "{Polarized ZZ pairs in gluon fusion and vector boson fusion at the LHC}",
    eprint = "2401.17365",
    archivePrefix = "arXiv",
    primaryClass = "hep-ph",
    reportNumber = "IFJPAN-IV-2024-2, COMETA-2024-01",
    month = "1",
    year = "2024"
}

@article{Hoppe:2023uux,
    author = {Hoppe, Mareen and Sch\"onherr, Marek and Siegert, Frank},
    title = "{Polarised cross sections for vector boson production with Sherpa}",
    eprint = "2310.14803",
    archivePrefix = "arXiv",
    primaryClass = "hep-ph",
    reportNumber = "IPPP/23/60, MCNET-23-11",
    doi = "10.1007/JHEP04(2024)001",
    journal = "JHEP",
    volume = "04",
    pages = "001",
    year = "2024"
}

@article{Pelliccioli:2023zpd,
    author = "Pelliccioli, Giovanni and Zanderighi, Giulia",
    title = "{Polarised-boson pairs at the LHC with NLOPS accuracy}",
    eprint = "2311.05220",
    archivePrefix = "arXiv",
    primaryClass = "hep-ph",
    doi = "10.1140/epjc/s10052-023-12347-4",
    journal = "Eur. Phys. J. C",
    volume = "84",
    number = "1",
    pages = "16",
    year = "2024"
}

@article{Denner:2026phn,
    author = "Denner, A. and Lombardi, D. and Lopez Portillo Chavez, S. and Pellen, M. and Pelliccioli, G.",
    title = "{MoCaNLO: a Monte Carlo integrator for NLO calculations}",
    eprint = "2602.19842",
    archivePrefix = "arXiv",
    primaryClass = "hep-ph",
    reportNumber = "COMETA-2026-03, FR-PHENO-26-04",
    month = "2",
    year = "2026"
}

@article{Denner:2020eck,
    author = "Denner, Ansgar and Pelliccioli, Giovanni",
    title = "{NLO QCD predictions for doubly-polarized WZ production at the LHC}",
    eprint = "2010.07149",
    archivePrefix = "arXiv",
    primaryClass = "hep-ph",
    doi = "10.1016/j.physletb.2021.136107",
    journal = "Phys. Lett. B",
    volume = "814",
    pages = "136107",
    year = "2021"
}

@article{Denner:2020bcz,
    author = "Denner, Ansgar and Pelliccioli, Giovanni",
    title = "{Polarized electroweak bosons in ${\bf \text{W}^+\text{W}^-}$ production at the LHC including NLO QCD effects}",
    eprint = "2006.14867",
    archivePrefix = "arXiv",
    primaryClass = "hep-ph",
    doi = "10.1007/JHEP09(2020)164",
    journal = "JHEP",
    volume = "09",
    pages = "164",
    year = "2020"
}

@article{Denner:2021csi,
    author = "Denner, Ansgar and Pelliccioli, Giovanni",
    title = "{NLO EW and QCD corrections to polarized ZZ production in the four-charged-lepton channel at the LHC}",
    eprint = "2107.06579",
    archivePrefix = "arXiv",
    primaryClass = "hep-ph",
    doi = "10.1007/JHEP10(2021)097",
    journal = "JHEP",
    volume = "10",
    pages = "097",
    year = "2021"
}

@article{Denner:2023ehn,
    author = "Denner, Ansgar and Haitz, Christoph and Pelliccioli, Giovanni",
    title = "{NLO EW corrections to polarised W+W\ensuremath{-} production and decay at the LHC}",
    eprint = "2311.16031",
    archivePrefix = "arXiv",
    primaryClass = "hep-ph",
    reportNumber = "COMETA-2023-01, MPP-2023-268",
    doi = "10.1016/j.physletb.2024.138539",
    journal = "Phys. Lett. B",
    volume = "850",
    pages = "138539",
    year = "2024"
}

@article{Denner:2022riz,
    author = "Denner, Ansgar and Haitz, Christoph and Pelliccioli, Giovanni",
    title = "{NLO QCD corrections to polarized diboson production in semileptonic final states}",
    eprint = "2211.09040",
    archivePrefix = "arXiv",
    primaryClass = "hep-ph",
    doi = "10.1103/PhysRevD.107.053004",
    journal = "Phys. Rev. D",
    volume = "107",
    number = "5",
    pages = "053004",
    year = "2023"
}

@article{Dao:2023kwc,
    author = "Dao, Thi Nhung and Le, Duc Ninh",
    title = "{NLO electroweak corrections to doubly-polarized $W^+W^-$ production at the LHC}",
    eprint = "2311.17027",
    archivePrefix = "arXiv",
    primaryClass = "hep-ph",
    doi = "10.1140/epjc/s10052-024-12579-y",
    journal = "Eur. Phys. J. C",
    volume = "84",
    number = "3",
    pages = "244",
    year = "2024"
}

@article{Ballestrero:2020qgv,
    author = "Ballestrero, Alessandro and Maina, Ezio and Pelliccioli, Giovanni",
    title = "{Different polarization definitions in same-sign $WW$ scattering at the LHC}",
    eprint = "2007.07133",
    archivePrefix = "arXiv",
    primaryClass = "hep-ph",
    doi = "10.1016/j.physletb.2020.135856",
    journal = "Phys. Lett. B",
    volume = "811",
    pages = "135856",
    year = "2020"
}

@article{Ballestrero:2019qoy,
    author = "Ballestrero, Alessandro and Maina, Ezio and Pelliccioli, Giovanni",
    title = "{Polarized vector boson scattering in the fully leptonic WZ and ZZ channels at the LHC}",
    eprint = "1907.04722",
    archivePrefix = "arXiv",
    primaryClass = "hep-ph",
    reportNumber = "VBSCAN-PUB-06-19",
    doi = "10.1007/JHEP09(2019)087",
    journal = "JHEP",
    volume = "09",
    pages = "087",
    year = "2019"
}

@article{Poncelet:2021jmj,
    author = "Poncelet, Rene and Popescu, Andrei",
    title = "{NNLO QCD study of polarised W$^{+}$W$^{-}$ production at the LHC}",
    eprint = "2102.13583",
    archivePrefix = "arXiv",
    primaryClass = "hep-ph",
    doi = "10.1007/JHEP07(2021)023",
    journal = "JHEP",
    volume = "07",
    pages = "023",
    year = "2021"
}

@article{Denner:2024tlu,
    author = "Denner, Ansgar and Haitz, Christoph and Pelliccioli, Giovanni",
    title = "{NLO EW and QCD corrections to polarised same-sign WW scattering at the LHC}",
    eprint = "2409.03620",
    archivePrefix = "arXiv",
    primaryClass = "hep-ph",
    reportNumber = "COMETA-2024-22, MPP-2024-178",
    month = "9",
    year = "2024"
}

@article{Dao:2024ffg,
    author = "Dao, Thi Nhung and Le, Duc Ninh",
    title = "{Polarized $W^+W^-$ pairs at the LHC: Effects from bottom-quark induced processes at NLO QCD+EW}",
    eprint = "2409.06396",
    archivePrefix = "arXiv",
    primaryClass = "hep-ph",
    month = "9",
    year = "2024"
}

@article{Pellen:2021vpi,
    author = "Pellen, Mathieu and Poncelet, Rene and Popescu, Andrei",
    title = "{Polarised W+j production at the LHC: a study at NNLO QCD accuracy}",
    eprint = "2109.14336",
    archivePrefix = "arXiv",
    primaryClass = "hep-ph",
    reportNumber = "CAVENDISH-HEP-21/13, FR-PHENO-2021-11",
    doi = "10.1007/JHEP02(2022)160",
    journal = "JHEP",
    volume = "02",
    pages = "160",
    year = "2022"
}

@Article{Le:2022ppa,
  author        = {Le, Duc Ninh and Baglio, Julien and Dao, Thi Nhung},
  title         = {{Doubly-polarized WZ hadronic production at NLO QCD+EW: calculation method and further results}},
  journal       = {Eur. Phys. J. C},
  year          = {2022},
  volume        = {82},
  number        = {12},
  pages         = {1103},
  archiveprefix = {arXiv},
  doi           = {10.1140/epjc/s10052-022-11032-2},
  eprint        = {2208.09232},
  primaryclass  = {hep-ph},
  reportnumber  = {CERN-TH-2022-110},
}

@Article{Le:2022lrp,
  author        = {Le, Duc Ninh and Baglio, Julien},
  title         = {{Doubly-polarized WZ hadronic cross sections at NLO QCD + EW accuracy}},
  journal       = {Eur. Phys. J. C},
  year          = {2022},
  volume        = {82},
  number        = {10},
  pages         = {917},
  archiveprefix = {arXiv},
  doi           = {10.1140/epjc/s10052-022-10887-9},
  eprint        = {2203.01470},
  primaryclass  = {hep-ph},
  reportnumber  = {CERN-TH-2022-027},
}

@Article{Dao:2023pkl,
  author        = {Dao, Thi Nhung and Le, Duc Ninh},
  title         = {{Enhancing the doubly-longitudinal polarization in WZ production at the LHC}},
  journal       = {Commun. in Phys.},
  year          = {2023},
  volume        = {33},
  number        = {3},
  pages         = {223},
  archiveprefix = {arXiv},
  doi           = {10.15625/0868-3166/18077},
  eprint        = {2302.03324},
  primaryclass  = {hep-ph},
}

@article{Denner:2025xdz,
    author = "Denner, Ansgar and Franken, Robert and Haitz, Christoph and Lombardi, Daniele and Pelliccioli, Giovanni",
    title = "{Electroweak corrections to doubly polarised WZ scattering at the LHC}",
    eprint = "2510.26462",
    archivePrefix = "arXiv",
    primaryClass = "hep-ph",
    reportNumber = "COMETA-2025-49",
    doi = "10.1007/JHEP02(2026)120",
    journal = "JHEP",
    volume = "02",
    pages = "120",
    year = "2026"
}

@article{Pelliccioli:2025com,
    author = "Pelliccioli, Giovanni and Poncelet, Rene",
    title = "{Precise predictions for joint polarisation fractions in WZ production at the LHC}",
    eprint = "2510.25898",
    archivePrefix = "arXiv",
    primaryClass = "hep-ph",
    reportNumber = "COMETA-2025-50, IFJPAN-IV-2025-22",
    month = "10",
    year = "2025"
}

@article{ATLAS:2022oge,
    author = "Aad, Georges and others",
    collaboration = "ATLAS",
    title = "{Observation of gauge boson joint-polarisation states in W\ensuremath{\pm}Z production from pp collisions at s=13 TeV with the ATLAS detector}",
    eprint = "2211.09435",
    archivePrefix = "arXiv",
    primaryClass = "hep-ex",
    reportNumber = "CERN-EP-2022-202",
    doi = "10.1016/j.physletb.2023.137895",
    journal = "Phys. Lett. B",
    volume = "843",
    pages = "137895",
    year = "2023"
}

@article{ATLAS:2024qbd,
    author = "Aad, Georges and others",
    collaboration = "ATLAS",
    title = "{Studies of the Energy Dependence of Diboson Polarization Fractions and the Radiation-Amplitude-Zero Effect in WZ Production with the ATLAS Detector}",
    eprint = "2402.16365",
    archivePrefix = "arXiv",
    primaryClass = "hep-ex",
    reportNumber = "CERN-EP-2024-045",
    doi = "10.1103/PhysRevLett.133.101802",
    journal = "Phys. Rev. Lett.",
    volume = "133",
    number = "10",
    pages = "101802",
    year = "2024"
}

@article{ATLAS:2023zrv,
    author = "Aad, Georges and others",
    collaboration = "ATLAS",
    title = "{Evidence of pair production of longitudinally polarised vector bosons and study of CP properties in ZZ \textrightarrow{} 4\ensuremath{\ell} events with the ATLAS detector at $ \sqrt{s} $ = 13 TeV}",
    eprint = "2310.04350",
    archivePrefix = "arXiv",
    primaryClass = "hep-ex",
    reportNumber = "CERN-EP-2023-199",
    doi = "10.1007/JHEP12(2023)107",
    journal = "JHEP",
    volume = "12",
    pages = "107",
    year = "2023"
}

@article{CMS:2021icx,
    author = "Tumasyan, Armen and others",
    collaboration = "CMS",
    title = "{Measurement of the inclusive and differential WZ production cross sections, polarization angles, and triple gauge couplings in pp collisions at $ \sqrt{s} $ = 13 TeV}",
    eprint = "2110.11231",
    archivePrefix = "arXiv",
    primaryClass = "hep-ex",
    reportNumber = "CMS-SMP-20-014, CERN-EP-2021-163",
    doi = "10.1007/JHEP07(2022)032",
    journal = "JHEP",
    volume = "07",
    pages = "032",
    year = "2022"
}

@article{CMS:2020etf,
    author = "Sirunyan, Albert M and others",
    collaboration = "CMS",
    title = "{Measurements of production cross sections of polarized same-sign W boson pairs in association with two jets in proton-proton collisions at $\sqrt{s} =$ 13 TeV}",
    eprint = "2009.09429",
    archivePrefix = "arXiv",
    primaryClass = "hep-ex",
    reportNumber = "CMS-SMP-20-006, CERN-EP-2020-168",
    doi = "10.1016/j.physletb.2020.136018",
    journal = "Phys. Lett. B",
    volume = "812",
    pages = "136018",
    year = "2021"
}

@article{CMS:2011kaj,
    author = "Chatrchyan, Serguei and others",
    collaboration = "CMS",
    title = "{Measurement of the Polarization of W Bosons with Large Transverse Momenta in W+Jets Events at the LHC}",
    eprint = "1104.3829",
    archivePrefix = "arXiv",
    primaryClass = "hep-ex",
    reportNumber = "CERN-PH-EP-2011-043, CMS-EWK-10-014",
    doi = "10.1103/PhysRevLett.107.021802",
    journal = "Phys. Rev. Lett.",
    volume = "107",
    pages = "021802",
    year = "2011"
}

@article{ATLAS:2019bsc,
    author = "Aaboud, Morad and others",
    collaboration = "ATLAS",
    title = "{Measurement of $W^{\pm}Z$ production cross sections and gauge boson polarisation in $pp$ collisions at $\sqrt{s} = 13$ TeV with the ATLAS detector}",
    eprint = "1902.05759",
    archivePrefix = "arXiv",
    primaryClass = "hep-ex",
    reportNumber = "CERN-EP-2018-327",
    doi = "10.1140/epjc/s10052-019-7027-6",
    journal = "Eur. Phys. J. C",
    volume = "79",
    number = "6",
    pages = "535",
    year = "2019"
}

@article{ATLAS:2016fbc,
    author = "Aaboud, Morad and others",
    collaboration = "ATLAS",
    title = "{Measurement of the W boson polarisation in $t\bar{t}$ events from pp collisions at $\sqrt{s}$ = 8 TeV in the lepton + jets channel with ATLAS}",
    eprint = "1612.02577",
    archivePrefix = "arXiv",
    primaryClass = "hep-ex",
    reportNumber = "CERN-EP-2016-219, CERN-PH-2016-219",
    doi = "10.1140/epjc/s10052-017-4819-4",
    journal = "Eur. Phys. J. C",
    volume = "77",
    number = "4",
    pages = "264",
    year = "2017",
    note = "[Erratum: Eur.Phys.J.C 79, 19 (2019)]"
}

@article{CMS:2016asd,
    author = "Khachatryan, Vardan and others",
    collaboration = "CMS",
    title = "{Measurement of the W boson helicity fractions in the decays of top quark pairs to lepton $+$ jets final states produced in pp collisions at $\sqrt s=$ 8TeV}",
    eprint = "1605.09047",
    archivePrefix = "arXiv",
    primaryClass = "hep-ex",
    reportNumber = "CMS-TOP-13-008, CERN-EP-2016-118",
    doi = "10.1016/j.physletb.2016.10.007",
    journal = "Phys. Lett. B",
    volume = "762",
    pages = "512--534",
    year = "2016"
}

@article{CMS:2020ezf,
    author = "Aad, Georges and others",
    collaboration = "CMS, ATLAS",
    title = "{Combination of the W boson polarization measurements in top quark decays using ATLAS and CMS data at $\sqrt{s} =$ 8 TeV}",
    eprint = "2005.03799",
    archivePrefix = "arXiv",
    primaryClass = "hep-ex",
    reportNumber = "CMS-TOP-19-004, ATLAS-TOPQ-2018-02, CERN-EP-2020-012",
    doi = "10.1007/JHEP08(2020)051",
    journal = "JHEP",
    volume = "08",
    number = "08",
    pages = "051",
    year = "2020"
}

@article{ATLAS:2025wuw,
    author = "Aad, Georges and others",
    collaboration = "ATLAS",
    title = "{Evidence for longitudinally polarized $W$ bosons in the electroweak production of same-sign $W$ boson pairs in association with two jets in pp collisions at $\sqrt{s}=13$ TeV with the ATLAS detector}",
    eprint = "2503.11317",
    archivePrefix = "arXiv",
    primaryClass = "hep-ex",
    reportNumber = "CERN-EP-2025-048",
    month = "3",
    year = "2025"
}

@Article{ATLAS:2016rnf,
  author        = {Aad, Georges and others},
  title         = {{Measurement of the angular coefficients in $Z$-boson events using electron and muon pairs from data taken at $\sqrt{s}=8$ TeV with the ATLAS detector}},
  journal       = {JHEP},
  year          = {2016},
  volume        = {08},
  pages         = {159},
  archiveprefix = {arXiv},
  collaboration = {ATLAS},
  doi           = {10.1007/JHEP08(2016)159},
  eprint        = {1606.00689},
  primaryclass  = {hep-ex},
  reportnumber  = {CERN-EP-2016-087},
}

@Article{CMS:2015cyj,
  author        = {Khachatryan, Vardan and others},
  title         = {{Angular coefficients of Z bosons produced in pp collisions at $\sqrt{s}$ = 8 TeV and decaying to $\mu^+ \mu^-$ as a function of transverse momentum and rapidity}},
  journal       = {Phys. Lett. B},
  year          = {2015},
  volume        = {750},
  pages         = {154--175},
  archiveprefix = {arXiv},
  collaboration = {CMS},
  doi           = {10.1016/j.physletb.2015.08.061},
  eprint        = {1504.03512},
  primaryclass  = {hep-ex},
  reportnumber  = {CMS-SMP-13-010, CERN-PH-EP-2015-046},
}

@Article{CMS:2020cph,
  author        = {Sirunyan, Albert M and others},
  title         = {{Measurements of the $W$ boson rapidity, helicity, double-differential cross sections, and charge asymmetry in $pp$ collisions at $\sqrt {s}$ = 13 TeV}},
  journal       = {Phys. Rev. D},
  year          = {2020},
  volume        = {102},
  number        = {9},
  pages         = {092012},
  archiveprefix = {arXiv},
  collaboration = {CMS},
  doi           = {10.1103/PhysRevD.102.092012},
  eprint        = {2008.04174},
  primaryclass  = {hep-ex},
  reportnumber  = {CMS-SMP-18-012, CERN-EP-2020-116},
}

@Article{ATLAS:2012au,
  author        = {Aad, Georges and others},
  title         = {{Measurement of the polarisation of $W$ bosons produced with large transverse momentum in $pp$ collisions at $\sqrt{s}=7$ TeV with the ATLAS experiment}},
  journal       = {Eur. Phys. J. C},
  year          = {2012},
  volume        = {72},
  pages         = {2001},
  archiveprefix = {arXiv},
  collaboration = {ATLAS},
  doi           = {10.1140/epjc/s10052-012-2001-6},
  eprint        = {1203.2165},
  primaryclass  = {hep-ex},
  reportnumber  = {CERN-PH-EP-2012-016},
}

@article{ATLAS:2022wnf,
    author = "Aad, Georges and others",
    collaboration = "ATLAS",
    title = "{Measurements of $Z\gamma+$jets differential cross sections in $pp$ collisions at $\sqrt{s}=13$ TeV with the ATLAS detector}",
    eprint = "2212.07184",
    archivePrefix = "arXiv",
    primaryClass = "hep-ex",
    reportNumber = "CERN-EP-2022-240",
    doi = "10.1007/JHEP07(2023)072",
    journal = "JHEP",
    volume = "07",
    pages = "072",
    year = "2023"
}

@article{ATLAS:2019rob,
    author = "Aaboud, Morad and others",
    collaboration = "ATLAS",
    title = "{Measurement of fiducial and differential $W^+W^-$ production cross-sections at $\sqrt{s}=13$  TeV with the ATLAS detector}",
    eprint = "1905.04242",
    archivePrefix = "arXiv",
    primaryClass = "hep-ex",
    reportNumber = "CERN-EP-2019-055",
    doi = "10.1140/epjc/s10052-019-7371-6",
    journal = "Eur. Phys. J. C",
    volume = "79",
    number = "10",
    pages = "884",
    year = "2019"
}

@article{ATLAS:2024yzu,
    author = "Aad, Georges and others",
    collaboration = "ATLAS",
    title = "{Measurements of $WH$ and $ZH$ production with Higgs boson decays into bottom quarks and direct constraints on the charm Yukawa coupling in $13\,\mathrm{TeV}$$pp$ collisions with the ATLAS detector}",
    eprint = "2410.19611",
    archivePrefix = "arXiv",
    primaryClass = "hep-ex",
    reportNumber = "CERN-EP-2024-237",
    month = "10",
    year = "2024"
}

@article{Cacciari:2008gp,
    author = "Cacciari, Matteo and Salam, Gavin P. and Soyez, Gregory",
    title = "{The anti-$k_t$ jet clustering algorithm}",
    eprint = "0802.1189",
    archivePrefix = "arXiv",
    primaryClass = "hep-ph",
    reportNumber = "LPTHE-07-03",
    doi = "10.1088/1126-6708/2008/04/063",
    journal = "JHEP",
    volume = "04",
    pages = "063",
    year = "2008"
}

@article{Buckley:2014ana,
    author = {Buckley, Andy and Ferrando, James and Lloyd, Stephen and Nordstr\"om, Karl and Page, Ben and R\"ufenacht, Martin and Sch\"onherr, Marek and Watt, Graeme},
    title = "{LHAPDF6: parton density access in the LHC precision era}",
    eprint = "1412.7420",
    archivePrefix = "arXiv",
    primaryClass = "hep-ph",
    reportNumber = "GLAS-PPE-2014-05, MCNET-14-29, IPPP-14-111, DCPT-14-222",
    doi = "10.1140/epjc/s10052-015-3318-8",
    journal = "Eur. Phys. J. C",
    volume = "75",
    pages = "132",
    year = "2015"
}

@article{Denner:2006ic,
    author = "Denner, Ansgar and Dittmaier, S.",
    editor = "Blumlein, J. and Moch, S. and Riemann, T.",
    title = "{The Complex-mass scheme for perturbative calculations with unstable particles}",
    eprint = "hep-ph/0605312",
    archivePrefix = "arXiv",
    reportNumber = "MPP-2006-67, PSI-PR-06-08",
    doi = "10.1016/j.nuclphysbps.2006.09.025",
    journal = "Nucl. Phys. B Proc. Suppl.",
    volume = "160",
    pages = "22--26",
    year = "2006"
}

@article{Denner:2005fg,
    author = "Denner, Ansgar and Dittmaier, S. and Roth, M. and Wieders, L. H.",
    title = "{Electroweak corrections to charged-current e+ e- ---\ensuremath{>} 4 fermion processes: Technical details and further results}",
    eprint = "hep-ph/0505042",
    archivePrefix = "arXiv",
    reportNumber = "MPP-2005-23, PSI-PR-05-05",
    doi = "10.1016/j.nuclphysb.2011.09.001",
    journal = "Nucl. Phys. B",
    volume = "724",
    pages = "247--294",
    year = "2005",
    note = "[Erratum: Nucl.Phys.B 854, 504--507 (2012)]"
}

@article{Denner:1999gp,
    author = "Denner, Ansgar and Dittmaier, S. and Roth, M. and Wackeroth, D.",
    title = "{Predictions for all processes e+ e- ---\ensuremath{>} 4 fermions + gamma}",
    eprint = "hep-ph/9904472",
    archivePrefix = "arXiv",
    reportNumber = "BI-TP-99-10, PSI-PR-99-12",
    doi = "10.1016/S0550-3213(99)00437-X",
    journal = "Nucl. Phys. B",
    volume = "560",
    pages = "33--65",
    year = "1999"
}

@article{Denner:2000bj,
    author = "Denner, Ansgar and Dittmaier, S. and Roth, M. and Wackeroth, D.",
    title = "{Electroweak radiative corrections to e+ e- ---\ensuremath{>} W W ---\ensuremath{>} 4 fermions in double pole approximation: The RACOONWW approach}",
    eprint = "hep-ph/0006307",
    archivePrefix = "arXiv",
    reportNumber = "BI-TP-2000-06, LU-ITP-2000-003, PSI-PR-00-11, UR-1604",
    doi = "10.1016/S0550-3213(00)00511-3",
    journal = "Nucl. Phys. B",
    volume = "587",
    pages = "67--117",
    year = "2000"
}

@article{Bertone:2017bme,
    author = "Bertone, Valerio and Carrazza, Stefano and Hartland, Nathan P. and Rojo, Juan",
    collaboration = "NNPDF",
    title = "{Illuminating the photon content of the proton within a global PDF analysis}",
    eprint = "1712.07053",
    archivePrefix = "arXiv",
    primaryClass = "hep-ph",
    reportNumber = "Nikhef/2017-064, CERN-TH-2017-235, NIKHEF-2017-064",
    doi = "10.21468/SciPostPhys.5.1.008",
    journal = "SciPost Phys.",
    volume = "5",
    number = "1",
    pages = "008",
    year = "2018"
}

@article{Manohar:2017eqh,
    author = "Manohar, Aneesh V. and Nason, Paolo and Salam, Gavin P. and Zanderighi, Giulia",
    title = "{The Photon Content of the Proton}",
    eprint = "1708.01256",
    archivePrefix = "arXiv",
    primaryClass = "hep-ph",
    reportNumber = "CERN-TH-2017-141",
    doi = "10.1007/JHEP12(2017)046",
    journal = "JHEP",
    volume = "12",
    pages = "046",
    year = "2017"
}

@article{Manohar:2016nzj,
    author = "Manohar, Aneesh and Nason, Paolo and Salam, Gavin P. and Zanderighi, Giulia",
    title = "{How bright is the proton? A precise determination of the photon parton distribution function}",
    eprint = "1607.04266",
    archivePrefix = "arXiv",
    primaryClass = "hep-ph",
    reportNumber = "CERN-TH-2016-155",
    doi = "10.1103/PhysRevLett.117.242002",
    journal = "Phys. Rev. Lett.",
    volume = "117",
    number = "24",
    pages = "242002",
    year = "2016"
}

@article{Schumann:2007mg,
    author = "Schumann, Steffen and Krauss, Frank",
    title = "{A Parton shower algorithm based on Catani-Seymour dipole factorisation}",
    eprint = "0709.1027",
    archivePrefix = "arXiv",
    primaryClass = "hep-ph",
    reportNumber = "DCPT-07-86, IPPP-07-43",
    doi = "10.1088/1126-6708/2008/03/038",
    journal = "JHEP",
    volume = "03",
    pages = "038",
    year = "2008"
}

@article{Yennie:1961ad,
    author = "Yennie, D. R. and Frautschi, Steven C. and Suura, H.",
    title = "{The infrared divergence phenomena and high-energy processes}",
    doi = "10.1016/0003-4916(61)90151-8",
    journal = "Annals Phys.",
    volume = "13",
    pages = "379--452",
    year = "1961"
}

@article{Schonherr:2008av,
    author = "Schonherr, Marek and Krauss, Frank",
    title = "{Soft Photon Radiation in Particle Decays in SHERPA}",
    eprint = "0810.5071",
    archivePrefix = "arXiv",
    primaryClass = "hep-ph",
    reportNumber = "DCPT-07-96, IPPP-07-48, MCNET-08-13",
    doi = "10.1088/1126-6708/2008/12/018",
    journal = "JHEP",
    volume = "12",
    pages = "018",
    year = "2008"
}

@article{Flower:2022iew,
    author = "Flower, Lois and Schoenherr, Marek",
    title = "{Photon splitting corrections to soft-photon resummation}",
    eprint = "2210.07007",
    archivePrefix = "arXiv",
    primaryClass = "hep-ph",
    reportNumber = "IPPP/22/69, MCnet-22-19",
    doi = "10.1007/JHEP03(2023)238",
    journal = "JHEP",
    volume = "03",
    pages = "238",
    year = "2023"
}

@article{Bierlich:2024vqo,
    author = "Bierlich, Christian and Buckley, Andy and Butterworth, Jonathan Mark and Gutschow, Christian and Lonnblad, Leif and Procter, Tomasz and Richardson, Peter and Yeh, Yoran",
    title = "{Robust independent validation of experiment and theory: Rivet version 4 release note}",
    eprint = "2404.15984",
    archivePrefix = "arXiv",
    primaryClass = "hep-ph",
    reportNumber = "MCNET-24-05",
    doi = "10.21468/SciPostPhysCodeb.36",
    journal = "SciPost Phys. Codeb.",
    volume = "36",
    pages = "1",
    year = "2024"
}

@Article{Carrivale:2025mjy,
  author        = {Carrivale, Costanza and others},
  title         = {{Precise Standard-Model predictions for polarised Z-boson pair production and decay at the LHC}},
  year          = {2025},
  month         = {5},
  archiveprefix = {arXiv},
  eprint        = {2505.09686},
  primaryclass  = {hep-ph},
  reportnumber  = {COMETA-2025-16, IFJPAN-IV-2025-10, IPPP/25/22, MCNET-25-07},
}

@Article{Haisch:2025jqr,
  author        = {Haisch, Ulrich and Linder, Jakob and Pelliccioli, Giovanni and Re, Emanuele and Zanderighi, Giulia},
  title         = {{Polarized-boson pairs at NLO in the SMEFT}},
  year          = {2025},
  month         = {7},
  archiveprefix = {arXiv},
  eprint        = {2507.21768},
  primaryclass  = {hep-ph},
  reportnumber  = {COMETA-2025-27, MPP-2025-136, LAPTH-027/25},
}

@Article{Hoche:2014kca,
  author        = {H{\"o}che, Stefan and Kuttimalai, Silvan and Schumann, Steffen and Siegert, Frank},
  title         = {{Beyond Standard Model calculations with Sherpa}},
  journal       = {Eur. Phys. J. C},
  year          = {2015},
  volume        = {75},
  number        = {3},
  pages         = {135},
  archiveprefix = {arXiv},
  doi           = {10.1140/epjc/s10052-015-3338-4},
  eprint        = {1412.6478},
  primaryclass  = {hep-ph},
  reportnumber  = {SLAC-PUB-16170, IPPP-14-105, DCPT-14-210, MCNET-14-35},
}

@Article{Richardson:2001df,
  author        = {Richardson, Peter},
  title         = {{Spin correlations in Monte Carlo simulations}},
  journal       = {JHEP},
  year          = {2001},
  volume        = {11},
  pages         = {029},
  archiveprefix = {arXiv},
  doi           = {10.1088/1126-6708/2001/11/029},
  eprint        = {hep-ph/0110108},
  reportnumber  = {CAVENDISH-HEP-2001-13, DAMTP-2001-83},
}

@Article{Knowles:1988vs,
  author       = {Knowles, I. G.},
  title        = {{Spin Correlations in Parton - Parton Scattering}},
  journal      = {Nucl. Phys. B},
  year         = {1988},
  volume       = {310},
  pages        = {571--588},
  doi          = {10.1016/0550-3213(88)90092-2},
  reportnumber = {Cavendish-HEP-88/5, NSF-ITP-88-36},
}

@Article{Knowles:1987cu,
  author       = {Knowles, I. G.},
  title        = {{Angular Correlations in {QCD}}},
  journal      = {Nucl. Phys. B},
  year         = {1988},
  volume       = {304},
  pages        = {767--793},
  doi          = {10.1016/0550-3213(88)90653-0},
  reportnumber = {CAVENDISH-HEP-87/5},
}

@Article{Collins:1987cp,
  author       = {Collins, John C.},
  title        = {{Spin Correlations in Monte Carlo Event Generators}},
  journal      = {Nucl. Phys. B},
  year         = {1988},
  volume       = {304},
  pages        = {794--804},
  doi          = {10.1016/0550-3213(88)90654-2},
  reportnumber = {IIT-87/11},
}

@Article{Gleisberg:2008fv,
  author        = {Gleisberg, Tanju and Hoeche, Stefan},
  title         = {{Comix, a new matrix element generator}},
  journal       = {JHEP},
  year          = {2008},
  volume        = {12},
  pages         = {039},
  archiveprefix = {arXiv},
  doi           = {10.1088/1126-6708/2008/12/039},
  eprint        = {0808.3674},
  primaryclass  = {hep-ph},
  reportnumber  = {SLAC-PUB-13232, IPPP-08-31, DCPT-08-62, MCNET-08-08},
}

@Article{Actis:2012qn,
  author        = {Actis, S. and Denner, A. and Hofer, L. and Scharf, A. and Uccirati, S.},
  title         = {{Recursive generation of one-loop amplitudes in the Standard Model}},
  journal       = {JHEP},
  year          = {2013},
  volume        = {04},
  pages         = {037},
  archiveprefix = {arXiv},
  doi           = {10.1007/JHEP04(2013)037},
  eprint        = {1211.6316},
  primaryclass  = {hep-ph},
  reportnumber  = {PSI-PR-12-09},
}

@Article{Actis:2016mpe,
  author        = {Actis, Stefano and Denner, Ansgar and Hofer, Lars and Lang, Jean-Nicolas and Scharf, Andreas and Uccirati, Sandro},
  title         = {{RECOLA: REcursive Computation of One-Loop Amplitudes}},
  journal       = {Comput. Phys. Commun.},
  year          = {2017},
  volume        = {214},
  pages         = {140--173},
  archiveprefix = {arXiv},
  doi           = {10.1016/j.cpc.2017.01.004},
  eprint        = {1605.01090},
  primaryclass  = {hep-ph},
  reportnumber  = {ICCUB-16-020},
}

@Article{Cascioli:2011va,
  author        = {Cascioli, Fabio and Maierhofer, Philipp and Pozzorini, Stefano},
  title         = {{Scattering Amplitudes with Open Loops}},
  journal       = {Phys. Rev. Lett.},
  year          = {2012},
  volume        = {108},
  pages         = {111601},
  archiveprefix = {arXiv},
  doi           = {10.1103/PhysRevLett.108.111601},
  eprint        = {1111.5206},
  primaryclass  = {hep-ph},
  reportnumber  = {ZU-TH-23-11, LPN11-66},
}

@Article{Buccioni:2019sur,
  author        = {Buccioni, Federico and Lang, Jean-Nicolas and Lindert, Jonas M. and Maierh{\"o}fer, Philipp and Pozzorini, Stefano and Zhang, Hantian and Zoller, Max F.},
  title         = {{OpenLoops 2}},
  journal       = {Eur. Phys. J. C},
  year          = {2019},
  volume        = {79},
  number        = {10},
  pages         = {866},
  archiveprefix = {arXiv},
  doi           = {10.1140/epjc/s10052-019-7306-2},
  eprint        = {1907.13071},
  primaryclass  = {hep-ph},
  reportnumber  = {IPPP/19/62, FR-PHENO-2019-12, PSI-PR-19-15, ZU-TH 37/19},
}

@Article{Le:2026oux,
  author        = {Le, Van Cuong and Le, Duc Ninh and Dao, Thi Nhung},
  title         = {{Triply polarized $WWW$ at the LHC: first glimpse at LO}},
  year          = {2026},
  month         = {3},
  archiveprefix = {arXiv},
  eprint        = {2603.02917},
  primaryclass  = {hep-ph},
}

@Article{ATLAS:2026kzu,
  author        = {Aad, Georges and others},
  title         = {{Evidence for longitudinally polarised $Z$ bosons in electroweak $W^{\pm}Z$ production in association with two jets from $pp$ collisions at $\sqrt{s} = $ 13 TeV and 13.6 TeV with the ATLAS detector}},
  year          = {2026},
  month         = {7},
  archiveprefix = {arXiv},
  collaboration = {ATLAS},
  eprint        = {2607.20246},
  primaryclass  = {hep-ex},
  reportnumber  = {CERN-EP-2026-213},
}

@Article{ATLAS:2026fhf,
  author        = {Aad, Georges and others},
  title         = {{Observation of pair production of longitudinally polarized $Z$ bosons in four-lepton final state with the ATLAS detector}},
  year          = {2026},
  month         = {7},
  archiveprefix = {arXiv},
  collaboration = {ATLAS},
  eprint        = {2607.12879},
  primaryclass  = {hep-ex},
  reportnumber  = {CERN-EP-2026-191},
}

@Article{Denner:2026vmf,
  author        = {Denner, Ansgar and Franken, Robert and Lopez Portillo Chavez, Santiago and Lombardi, Daniele and Pelliccioli, Giovanni},
  title         = {{The cleanest of them all: NLO electroweak corrections to vector-boson scattering into doubly polarised ZZ pairs at the LHC}},
  year          = {2026},
  month         = {7},
  archiveprefix = {arXiv},
  eprint        = {2607.01174},
  primaryclass  = {hep-ph},
  reportnumber  = {COMETA-2026-21},
}

@Article{Dittmaier:1998nn,
  author        = {Dittmaier, Stefan},
  title         = {{Weyl-van der Waerden formalism for helicity amplitudes of massive particles}},
  journal       = {Phys. Rev. D},
  year          = {1998},
  volume        = {59},
  pages         = {016007},
  archiveprefix = {arXiv},
  doi           = {10.1103/PhysRevD.59.016007},
  eprint        = {hep-ph/9805445},
  reportnumber  = {CERN-TH-98-143},
}

@Article{Krauss:2001iv,
  author        = {Krauss, F. and Kuhn, R. and Soff, G.},
  title         = {{AMEGIC++ 1.0: A Matrix element generator in C++}},
  journal       = {JHEP},
  year          = {2002},
  volume        = {02},
  pages         = {044},
  archiveprefix = {arXiv},
  doi           = {10.1088/1126-6708/2002/02/044},
  eprint        = {hep-ph/0109036},
  reportnumber  = {CAVENDISH-HEP-01-11},
}

@Article{Biedermann:2017yoi,
  author        = {Biedermann, Benedikt and Br{\"a}uer, Stephan and Denner, Ansgar and Pellen, Mathieu and Schumann, Steffen and Thompson, Jennifer M.},
  title         = {{Automation of NLO QCD and EW corrections with Sherpa and Recola}},
  journal       = {Eur. Phys. J. C},
  year          = {2017},
  volume        = {77},
  pages         = {492},
  archiveprefix = {arXiv},
  doi           = {10.1140/epjc/s10052-017-5054-8},
  eprint        = {1704.05783},
  primaryclass  = {hep-ph},
}

\end{document}